\documentclass[twocolumn]{aastex631}

\usepackage{newunicodechar}
\DeclareRobustCommand{\okina}{%
  \raisebox{\dimexpr\fontcharht\font`A-\height}{%
    \scalebox{0.8}{`}%
  }%
}
\newunicodechar{ʻ}{\okina}

\usepackage{multirow}
\usepackage{hyperref}
\usepackage{makecell}
\usepackage{gensymb}
\usepackage{amsmath}
\usepackage{stackengine}
\usepackage{tabularx}
\usepackage{threeparttable}
\usepackage{CJK}

\definecolor{DarkOrange}{RGB}{204, 85, 0}
\definecolor{LincolnGreen}{RGB}{17, 102, 0}
\definecolor{Rust}{HTML}{9B4F0F}
\definecolor{DarkCyan}{HTML}{008B8B}
\definecolor{MediumAquaMarine}{HTML}{66CDAA}
\definecolor{Maroon}{HTML}{800000}
\definecolor{Crimson}{HTML}{DC143C}

\defcitealias{Bruch+2023}{B23}
\defcitealias{Jacobson-Galan+2024_FMII}{JG24}

\renewcommand\labelenumi{(\roman{enumi})}
\renewcommand\theenumi\labelenumi

\newcommand{\BTSbot}{{\texttt{BTSbot} }}
\newcommand{\BTSbotnearby}{{\texttt{BTSbot-nearby} }}

\newcommand{\DM}{{31.33}}
\newcommand{\DMunc}{{0.43}}

\newcommand{\dist}{{18.45}}
\newcommand{\distunc}{{3.7}}

\newcommand{\z}{{0.00592}}
\newcommand{\zunc}{{0.00002}}

\newcommand{\AVhost}{{0.636}}
\newcommand{\AVhostunc}{{0.122}}

\newcommand{\AVMW}{{0.120}}

\newcommand{\NaIDEW}{{0.815}}
\newcommand{\NaIDEWunc}{{0.033}}

\shorttitle{SN\,2024jlf \& BTSbot-nearby}
\shortauthors{Rehemtulla et al.}

\begin{document}
\begin{CJK*}{UTF8}{gbsn}
\title{The \texttt{BTSbot-nearby} discovery of SN 2024jlf: rapid, autonomous follow-up\\probes interaction in an 18.5 Mpc Type IIP supernova}

\correspondingauthor{Nabeel Rehemtulla}
\email{nabeelr@u.northwestern.edu}

\author[0000-0002-5683-2389]{Nabeel~Rehemtulla}
\affiliation{Department of Physics and Astronomy, Northwestern University, 2145 Sheridan Road, Evanston, IL 60208, USA}
\affiliation{Center for Interdisciplinary Exploration and Research in Astrophysics (CIERA), 1800 Sherman Ave., Evanston, IL 60201, USA}
\affiliation{NSF-Simons AI Institute for the Sky (SkAI), 172 E. Chestnut St., Chicago, IL 60611, USA}

\author[0000-0002-3934-2644]{W.~V.~Jacobson-Gal\'{a}n}
\altaffiliation{NASA Hubble Fellow}
\affil{Department of Astronomy and Astrophysics, California Institute of Technology, Pasadena, CA 91125, USA}

\author[0000-0003-2091-622X]{Avinash~Singh}
\affil{Department of Astronomy, The Oskar Klein Center, Stockholm University, AlbaNova, SE-10691 Stockholm, Sweden}

\author[0000-0001-9515-478X]{Adam~A.~Miller}
\affiliation{Department of Physics and Astronomy, Northwestern University, 2145 Sheridan Road, Evanston, IL 60208, USA}
\affiliation{Center for Interdisciplinary Exploration and Research in Astrophysics (CIERA), 1800 Sherman Ave., Evanston, IL 60201, USA}
\affiliation{NSF-Simons AI Institute for the Sky (SkAI), 172 E. Chestnut St., Chicago, IL 60611, USA}

\author[0000-0002-5740-7747]{Charles~D.~Kilpatrick}
\affil{Center for Interdisciplinary Exploration and Research in Astrophysics (CIERA), 1800 Sherman Ave., Evanston, IL 60201, USA}

\author[0000-0002-0129-806X]{K-Ryan~Hinds}
\affil{Astrophysics Research Institute, Liverpool John Moores University, IC2, Liverpool Science Park, 146 Brownlow Hill, Liverpool L3 5RF, UK}

\author[0000-0002-7866-4531]{Chang~Liu (刘畅)}
\affil{Department of Physics and Astronomy, Northwestern University, 2145 Sheridan Road, Evanston, IL 60208, USA}
\affil{Center for Interdisciplinary Exploration and Research in Astrophysics (CIERA), 1800 Sherman Ave., Evanston, IL 60201, USA}

\author[0000-0001-6797-1889]{Steve~Schulze}
\affil{Center for Interdisciplinary Exploration and Research in Astrophysics (CIERA), 1800 Sherman Ave., Evanston, IL 60201, USA}

\author[0000-0003-1546-6615]{Jesper Sollerman}
\affiliation{Department of Astronomy, The Oskar Klein Center, Stockholm University, AlbaNova, SE-10691 Stockholm, Sweden}

\author[0009-0003-6181-4526]{Theophile~Jegou~du~Laz}
\affiliation{Division of Physics, Mathematics, and Astronomy, California Institute of Technology, Pasadena, CA 91125, USA}

\author[0000-0002-2184-6430]{Tom\'{a}s~Ahumada}
\affiliation{Division of Physics, Mathematics, and Astronomy, California Institute of Technology, Pasadena, CA 91125, USA}

\author[0000-0002-4449-9152]{Katie~Auchettl}
\affiliation{Department of Astronomy and Astrophysics, University of California, Santa Cruz, CA 95064, USA}
\affiliation{OzGrav, School of Physics, The University of Melbourne, VIC 3010, Australia}

\author[0000-0003-1325-6235]{S.~J.~Brennan}
\affiliation{Department of Astronomy, The Oskar Klein Center, Stockholm University, AlbaNova, SE-10691 Stockholm, Sweden}

\author[0000-0002-8262-2924]{Michael W. Coughlin} 
\affiliation{School of Physics and Astronomy, University of Minnesota, Minneapolis, MN 55455, USA}

\author[0000-0002-4223-103X]{Christoffer Fremling}
\affiliation{Division of Physics, Mathematics, and Astronomy, California Institute of Technology, Pasadena, CA 91125, USA}
\affiliation{Caltech Optical Observatories, California Institute of Technology, Pasadena, CA 91125, USA}

\author[0000-0002-3884-5637]{Anjasha Gangopadhyay}
\affil{Department of Astronomy, The Oskar Klein Center, Stockholm University, AlbaNova, SE-10691 Stockholm, Sweden}

\author[0000-0001-8472-1996]{Daniel A. Perley}
\affiliation{Astrophysics Research Institute, Liverpool John Moores University, IC2, Liverpool Science Park, 146 Brownlow Hill, Liverpool L3 5RF, UK}

\author[0000-0001-5847-7934]{Nikolaus Z.\ Prusinski}
\affiliation{Cahill Center for Astronomy and Astrophysics, California Institute of Technology, MC 249-17, Pasadena, CA 91125, USA}

\author[0000-0003-1227-3738]{Josiah Purdum}
\affiliation{Caltech Optical Observatories, California Institute of Technology, Pasadena, CA 91125, USA}

\author[0000-0003-3658-6026]{Yu-Jing Qin}
\affiliation{Division of Physics, Mathematics, and Astronomy, California Institute of Technology, Pasadena, CA 91125, USA}

\author[0009-0003-8153-9576]{Sara~Romagnoli}
\affiliation{OzGrav, School of Physics, The University of Melbourne, VIC 3010, Australia}

\author[0009-0008-3724-1824]{Jennifer~Shi}
\affiliation{OzGrav, School of Physics, The University of Melbourne, VIC 3010, Australia}

\author[0000-0003-0733-2916]{Jacob~L.~Wise}
\affiliation{Astrophysics Research Institute, Liverpool John Moores University, IC2, Liverpool Science Park, 146 Brownlow Hill, Liverpool L3 5RF, UK}

\author[0000-0001-9152-6224]{Tracy~X.~Chen}
\affiliation{IPAC, California Institute of Technology, 1200 E. California Blvd, Pasadena, CA 91125, USA}

\author[0000-0001-5668-3507]{Steven~L.~Groom}
\affiliation{IPAC, California Institute of Technology, 1200 E. California Blvd, Pasadena, CA 91125, USA}

\author[0000-0002-6230-0151]{David~O.~Jones}
\affiliation{Institute for Astronomy, University of Hawai\okina i, 640 N.~A\okina ohoku Pl., Hilo, HI 96720, USA}

\author[0000-0002-5619-4938]{Mansi M. Kasliwal}
\affiliation{Division of Physics, Mathematics, and Astronomy, California Institute of Technology, Pasadena, CA 91125, USA}

\author[0000-0001-7062-9726]{Roger Smith}
\affiliation{Caltech Optical Observatories, California Institute of Technology, Pasadena, CA 91125, USA}

\author{Niharika~Sravan}
\affiliation{Department of Physics, Drexel University, Philadelphia, PA 19104, USA}

\author[0000-0001-5390-8563]{Shrinivas R. Kulkarni}
\affiliation{Division of Physics, Mathematics, and Astronomy, California Institute of Technology, Pasadena, CA 91125, USA}



\begin{abstract}

We present observations of the Type IIP supernova (SN) 2024jlf, including spectroscopy beginning just 0.7 days ($\sim$17 hours) after first light. Rapid follow-up was enabled by the new \texttt{BTSbot-nearby} program, which involves autonomously triggering target-of-opportunity requests for new transients in Zwicky Transient Facility data that are coincident with nearby ($D<60$ Mpc) galaxies and identified by the \texttt{BTSbot} machine learning model. Early photometry and non-detections shortly prior to first light show that SN\,2024jlf initially brightened by $>$4 mag/day, quicker than $\sim$90\% of Type II SNe. Early spectra reveal weak flash ionization features: narrow, short-lived ($1.3 < \tau ~\mathrm{[d]} < 1.8$) emission lines of H$\alpha$, \ion{He}{2}, and \ion{C}{4}. Assuming a wind velocity of $v_w=50$ km s$^{-1}$, these properties indicate that the red supergiant progenitor exhibited enhanced mass-loss in the last year before explosion. We constrain the mass-loss rate to $10^{-4} < \dot{M}~\mathrm{[M_\odot~yr^{-1}]} < 10^{-3}$ by matching observations to model grids from two independent radiative hydrodynamics codes. \texttt{BTSbot-nearby} automation minimizes spectroscopic follow-up latency, enabling the observation of ephemeral early-time phenomena exhibited by transients.



%

\end{abstract}



\section{Introduction} \label{sec:intro}

Following shock breakout in core-collapse supernovae (CCSNe), the shock front propagates outwards and photoionizes the surrounding medium. Circumstellar material (CSM) in the vicinity will also be photoionized as the shock progresses, and, should the CSM be at high enough density, the subsequent recombination will produce ``flash," or ``IIn-like," features \citep{Gal-Yam+2014}: narrow ($v\lesssim100$ km s$^{-1}$), short-lived ($\sim$days) emission lines of highly ionized species (e.g., He~\textsc{ii}, C~\textsc{iii}/\textsc{iv}, N~\textsc{iii}/\textsc{iv}/\textsc{v}, O~\textsc{vi}). CSM with lower density (i.e. that arising from a progenitor with mass-loss rate $\dot{M}\lesssim10^{-5}~\textrm{M}_\odot~\textrm{yr}^{-1}$) will recombine within hours after shock breakout and not provide significant forward shock luminosity.

The properties of an event's flash features are directly linked to physical properties, namely, the CSM extent and mass and the chemical composition of the progenitor star's surface prior to the SN. Critically, properties of flash features also imply the progenitor star underwent an episode of enhanced mass-loss shortly before the SN: the high ionization features imply the CSM is dense and the short duration suggests the CSM is close-in to the progenitor star. Comparison to sophisticated radiative hydrodynamic models \citep[e.g., those from][]{Dessart+2017, BoianGroh2019, Moriya+2023, Dessart_WJG_2023} enable the inference of these parameters \citep[e.g.,][]{Shivvers+2015, BoianGroh2020, Zhang+2020, Tartaglia+2021, Terreran+2022, Jacobson-Galan+2022, Andrews+2024}. Progenitor mass-loss rates in the years before explosion have been inferred to upwards of $10^{-2}~\textrm{M}_\odot~\textrm{yr}^{-1}$ for some ``flashing" SNe (e.g., SN\,2023ixf \citealt{Bostroem+2023, Jacobson-Galan+2023, Teja+2023, Zimmerman+2023, Berger+2023, Hiramatsu+2023, Singh+2024}; and SN\,2024ggi \citealt{Jacobson-Galan+2024_24ggi, Shrestha+2024}). These $\dot{M}$ in the years before explosion are orders of magnitude higher than that of Milky Way red supergiants (RSGs) \citep[$\dot{M}\approx10^{-6}~\textrm{M}_\odot~\textrm{yr}^{-1}$;][]{Beasor+2020} and that inferred for Type IIP SNe without flash features (e.g., SN\,2017eaw \citealt{Kilpatrick+2018, Szalai+2019}).

Flash features have historically been difficult to study, in large part due to their ephemeral nature. \cite{Niemela+1985} published the first SN spectra exhibiting flash features, and \cite{Yaron+2017} published a remarkable dataset on SN\,2013fs, now a prototypical example of flashing Type II SNe (SNe\,II). The search for these features (``flash spectroscopy") has produced samples of SNe\,II with spectra shortly after first light, including dozens of events exhibiting flash features \citep{Khazov+2016, Bruch+2021, Bruch+2023, Jacobson-Galan+2024_FMII}. Short-lived narrow emission features have also been seen in SNe from stripped stars \citep[e.g.,][]{Gangopadhyay+2020, Gangopadhyay+2022, Perley+2022, Davis+2023, Schulze+2024}, but we discuss only SNe\,II here.


\cite{Bruch+2023} (\citetalias{Bruch+2023} henceforth; preceded by \citealt{Bruch+2021} and accompanied by \citealt{Irani+2024}) search for narrow \ion{He}{2} $\lambda4686$ emission in SNe\,II, which they use to define a SN as flashing. They unambiguously identify such emission in at least 30\% of events in their sample while more than half have moderately confident detections. They also find no significant difference in the optical light curve properties (rise time, peak magnitude, color at peak) of flashing and non-flashing events while \cite{Irani+2024} find that flashers have larger initial blackbody radii and luminosities. Crucially, \cite{Irani+2024} find that many SNe\,II do not fit shock cooling models well in the early time but do later on in their evolution, underscoring the importance of well-sampled early optical and UV light curves. \citetalias{Bruch+2023} quantify the duration of flash features ($\tau$) as the duration for which narrow \ion{He}{2} $\lambda4686$ emission is visible. The events in their sample typically have $\tau\gtrsim5$ days, and they find tentative evidence for an additional population of rare events with long-lived flash features ($\tau>10$ days). 

\cite{Jacobson-Galan+2024_FMII} (\citetalias{Jacobson-Galan+2024_FMII} henceforth)  focus on features matching slightly different criteria which they call ``IIn-like:" symmetric, narrow, short-lived features with Lorentzian wings. The Lorentzian wings are caused by electron scattering before photons escape from the CSM.\footnote{Hence the name IIn-like as this is similar to the process which forms the narrow features present in Type IIn SNe.} Which species are seen exhibiting IIn-like features and their relative strengths determines which of three classes \citetalias{Jacobson-Galan+2024_FMII} classifies a SN\,II as. They find that each of these classes exhibits a different distribution of IIn-like feature duration ($t_{\mathrm{IIn}}$), and that there are clear differences in light curve properties (peak optical and UV magnitude and pseudo-bolometric luminosity) between flashing and non-flashing events.

Still, there remain large uncertainties in the results of these sample studies, primarily due to the relatively small sample sizes and the lack of systematic follow-up observations.
Addressing these challenges necessitates observational campaigns that start sooner after explosion and continue with high-cadence, high signal-to-noise spectroscopy at least until the narrow features have subsided, an important transition for distinguishing flashing SNe\,II from Type IIn SNe.

The favored facilities for finding SNe shortly after explosion are wide-field time-domain surveys like ATLAS \citep[Asteroid Terrestrial Last-Alert System;][]{Tonry+2011, Tonry+2018, Smith+2020}, DLT40 \citep[Distance Less Than 40;][]{Tartaglia+2018}, YSE \citep[Young Supernova Experiment;][]{Jones+2021} on the Pan-STARRS telescopes \citep[Panoramic Survey Telescope and Rapid Response System;][]{Kaiser+2002}, and ZTF \citep[Zwicky Transient Facility;][]{Bellm+2019a, Bellm+2019b, Graham+2019, Dekany+2020, Masci+2019}. The early stages of the established real-time workflows in these surveys, e.g. data reduction, alert generation, and real/bogus classification \citep{Bloom+2012, Brink+2013}, are typically fully automated. SNe discovery, however, relies almost entirely on visual inspection of SN candidates (or ``scanning"). Because of this, humans are among the greatest sources of latency between a survey collecting data and follow-up spectroscopy being obtained for a new SN in that data. 

Latency in transient workflows can be minimized by automating the process of transient identification and follow-up. Doing so introduces two key challenges: (i) spectroscopic resources are extremely valuable so identifications following filtering must be made with very low false positive (FP) rates; (ii) truly minimizing latency requires making identifications with minimal data, i.e. often only two detections over one to two nights. Here, we focus on the classes of relatively common optical transients (predominantly SNe) and put aside more exotic classes like gamma-ray bursts and electromagnetic counterparts to gravitational wave events. Although the challenges are broadly similar, the multi-wavelength and multi-messenger nature of these events make the workflows quite different \citep[e.g.,][]{DElia+2009}. Many tools exist for automatically identifying and characterizing SN candidates, but few satisfy both criteria, making them unfit for autonomous low-latency SN follow-up. Photometric transient classification tools \citep[e.g.,][]{Boone+2019, Muthukrishna+2019, Villar+2019, Villar+2020, Gagliano+2023} often require days to weeks of data to perform well. Many other tools focus on rare types of transients and thus typically compromise purity to maximize the number of recovered events \citep[e.g.,][]{Gomez+2020, Gomez+2023, Stein+2024}. Tools which perform high-level classification of events (e.g., transient vs. other) on real-time alert streams, however, are better suited for this task \citep[e.g.,][]{Duev+2021, Carrasco-Davis+2021, Rehemtulla+2024_BTSbot}.

One such tool is \texttt{BTSbot} \citep{Rehemtulla+2024_BTSbot}, a machine learning (ML) model for autonomously identifying bright ($m_\mathrm{peak}\leq18.5\,\mathrm{mag}$) extragalactic transients in the ZTF data stream. \texttt{BTSbot} was designed to select targets for the ZTF Bright Transient Survey \citep[BTS;][]{Fremling+2020, Perley+2020, Rehemtulla+2024_BTSbot}, which endeavors to spectroscopically classify all such sources. \texttt{BTSbot} was deployed into ZTF and BTS operations in 2023 October with the ability to autonomously trigger follow-up to the SED Machine \citep[SEDM;][]{Blagorodnova+2018, Kim+2022}. Since then, it has sent $>$1200 automated spectroscopic follow-up requests, many of which are on transients it identified before human scanners in BTS. Close monitoring of triggers in the first eight months of operations revealed that $\sim$96\% of \BTSbot identifications were of genuine extragalactic transients, matching expected performance \citep{Rehemtulla+2024_BTSbot}. 

The discovery and follow-up of SN\,2024jlf demonstrated the efficacy of automating rapid response follow-up of quickly evolving transients. This was done with a new program repurposing the \BTSbot model: \texttt{BTSbot-nearby}. SN\,2024jlf was a Type IIP SN, the core-collapse explosion of a RSG star \citep[e.g.,][for review]{Hamuy2003, Pejcha+2015}. After reaching peak luminosity, these events exhibit a characteristic prolonged light curve plateau phase (encoded by the ``P" in ``SN\,IIP") where the SN remains roughly constant in luminosity for $\sim$100 days \citep{Chugai1991}. Spectroscopic observations during this phase reveal broad hydrogen Balmer P-Cygni features. 

Early spectroscopic observations of SN\,2024jlf revealed flash features in \ion{He}{2}, \ion{C}{4}, and H$\alpha$. We match observations of SN\,2024jlf to models produced with sophisticated radiation hydrodynamic simulations in order to infer properties of the progenitor star and CSM. We also discuss discrepancies in values inferred from models produced with different simulation codes.

Throughout this study we assume a flat $\Lambda$CDM cosmology with $H_0=70$ $\mathrm{km\,s^{-1}\,Mpc^{-1}}$ and $\Omega_M=0.3$. All times here are in UTC unless otherwise noted, and all magnitudes are in the AB system unless otherwise noted.

\section{BTSbot-nearby} \label{sec:btsbot_nearby}


We present \texttt{BTSbot-nearby}: a new autonomous, rapid follow-up program for nearby ($D<60$ Mpc) infant SNe in ZTF data. This program involves repurposing the \BTSbot model and introducing new filtering to identify nearby infant SNe with a very low FP rate.

The \BTSbotnearby alert filter\footnote{\BTSbotnearby was initially introduced in \cite{Rehemtulla+2024_nearby_AN}. Here, we describe the most up-to-date version of the filtering.} is based on that used by BTS and \texttt{BTSbot} (see \citealt{Perley+2020}) and the ZTF Census of the Local Universe experiment \cite[CLU;][]{De+2020}. 

We make some minor adjustments to criteria in these filters. The real/bogus score threshold using \texttt{braai} \citep{Duev+2019}, ZTF's deep learning model for filtering of non-astrophysical alerts, is increased from 0.3 (in the BTS alert fitler) to 0.7. This adjustment is made to more consistently reject bogus alerts which frequently occur atop very bright galactic nuclei. Such backgrounds which are common for nearby transients and alerts from these areas tend to be misclassified by \texttt{braai}. Based on Figure~9 of \cite{Duev+2019}, this should reduce the real/bogus false positive rate from $\sim$3.6\% to $\sim$1.6\%. We also make the rejection of candidates based on crossmatches to the Minor Planet Center (MPC) catalog more strict; a source with any alert within 2" of an MPC object will be rejected.


The primary addition is filtering following host galaxy association. Alerts only pass the \BTSbotnearby filter if they are coincident with a NASA Extragalactic Database Local Volume Sample \citep[NED-LVS;][]{Cook+2023, Cook+2025} galaxy that has distance $D<60$~Mpc. An alert is considered coincident with a galaxy if its projected physical offset is $\leq$15 kpc, capped at a maximum angular offset of $2'$. The angular offset limit is imposed because the size of the selection region for very nearby galaxies ($D\lesssim1$ Mpc) is otherwise extremely large and prone to selecting sources unrelated to the nearby galaxy.
The 60 Mpc limit is motivated by the ZTF sensitivity: a 20.5 mag source at 60 Mpc corresponds to having an absolute magnitude of roughly $-13.5$ mag, sufficient to capture SNe very early. 
An offset limit of 15 kpc selects the vast majority of CCSNe and SNe\,Ia \citep{Fremling+2020, Schulze+2021} while limiting contamination from sources projected nearby but unrelated to a galaxy.

Once a source is associated to a galaxy, the galaxy's distance can be used to estimate the source's absolute magnitude. \BTSbotnearby requires that the source have absolute magnitude in any ZTF filter of $M~<~-11$~mag to reject variable stars and classical novae which are visible to ZTF in the most nearby galaxies \citep[e.g., M31 and M33;][]{Capaccioli+1989}. The source must also have at least one alert with \BTSbot score greater than 0.5. This criterion is crucial for rejecting remaining contamination like non-SN stellar activity in very nearby galaxies, active galactic nuclei and cataclysmic variables projected over nearby galaxies, and more. The target must also not be coincident with a source on the Transient Name Server (TNS)\footnote{\url{https://www.wis-tns.org}} that was reported more than 36 hours before the time of filtering. Next, we select only sources with non-detections within 3.5 days of the latest alert to ensure that those transients which are selected are very young. The final criteria considers the status of the source on ZTF's first-party marshal Fritz\footnote{\url{https://github.com/fritz-marshal/fritz}} \citep[a SkyPortal instance;][]{van_der_Walt+2019, Coughlin+2023}; a trigger is prevented if there is (i) an existing spectroscopic follow-up request; (ii) a classification assigned; or (iii) a spectrum already present. The criteria involving the status of the source on Fritz are described in detail in Appendix~B of \cite{Rehemtulla+2024_BTSbot}.
Once all criteria are met, follow-up requests can be triggered with configurable parameters. 

\texttt{BTSbot-nearby} filtering runs on ZTF's first-party alert broker, \texttt{Kowalski} \citep{Duev+2019}, and marshal, Fritz, which support triggering automated follow-up to numerous other facilities. 

\subsection{Quality of \BTSbotnearby filtering relative to human scanning}

We characterize the quality of the \BTSbotnearby filtering by assessing the completeness (or ``recall") and purity (or ``precision") of triggers and comparing the latency of automated follow-up with that of traditional, human-triggered follow-up. We pass ZTF public and partnership alerts from the main surveys between 1 October 2023 to 1 April 2025\footnote{The dates are determined by when the latest version of \BTSbot was deployed into production and when this analysis was conducted.} through the \BTSbotnearby filters and log which sources satisfy them and whether they are true positives (TPs) or FPs. False negatives (FNs) are identified by cross-matching with sources cataloged in BTS or CLU during the same time frame.





\begin{figure}
    \begin{center}
    \includegraphics[width=1\columnwidth]{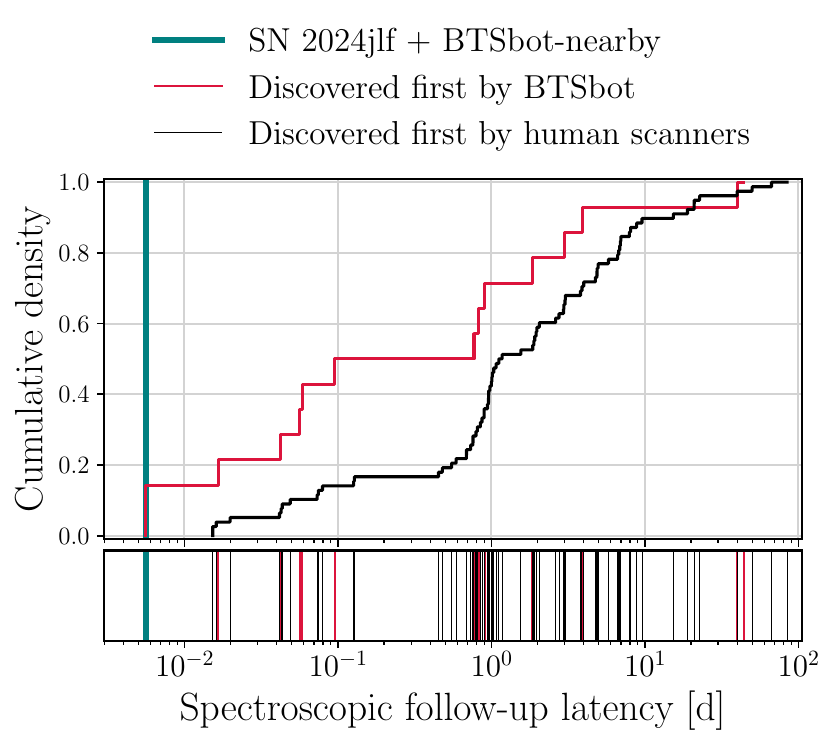}
    \caption{Comparison of spectroscopic follow-up latency ($\Delta t_{\mathrm{spec}}$) distributions for automated (red, teal) and manual (black) follow-up of nearby ($D<60$ Mpc) SNe discovered in ZTF data. $\Delta t_{\mathrm{spec}}$ is the time between a SN passing a ZTF alert filter and the first spectrum being taken. Relying on human-triggered follow-up often incurs $\Delta t_{\mathrm{spec}}\gtrsim1$ day of latency whereas automated follow-up can expedite $\Delta t_{\mathrm{spec}}$ by an order of magnitude. \BTSbotnearby demonstrated $\Delta t_{\mathrm{spec}}\approx7$ minutes for SN\,2024jlf.
    }
    \label{fig:latency}
    \end{center}
\end{figure}

Statistics are calculated identically to how they were by \cite{Rehemtulla+2024_BTSbot}.
77 sources satisfy the \BTSbotnearby filtering in this time frame; $\sim90\%$ of these are genuine nearby transients, identified via cross-match with BTS/CLU catalogs and visual inspection. The FPs are strongly dominated by distant transients that are erroneously cross-matched to nearby galaxies but also include small numbers of active galactic nuclei (AGN) and cataclysmic variable stars (CVs).  The count of FNs are definition dependent. In most cases, BTS/CLU transients were not selected by \BTSbotnearby filtering because their host galaxies do not appear in the NED catalog used. In some cases, the host galaxies had entries in NED but lacked either a spectroscopic redshift or a redshift-independent distance measurement. Transients which are very highly offset from their host represent a very small but astrophysically important population of FNs. These results suggest that that the host galaxy association mechanism used by \BTSbotnearby can still be improved.

To illustrate the minimization of follow-up latency with \texttt{BTSbot-nearby}, we define metrics to compare with the established filtering and follow-up in ZTF. We measure the ``spectroscopic follow-up latency" ($\Delta t_{\mathrm{spec}}$) as the time between when a source first passes the alert filter, i.e. when it is then available for humans to scan or \BTSbot to identify, and when the first spectrum of the source is taken. This metric is favorable because it is not sensitive to the ZTF observing strategy or to variations in processing time at IPAC, while also not limiting the analysis to a subset of spectrographs used. The time of a follow-up request being submitted is not used because it is only attainable for SEDM requests. We compute this metric for $D\lesssim60$ Mpc SNe between 1 January 2021 (around when the BTS alert filter was last improved) and 1 October 2024 (when this analysis was conducted). We also limit this analysis to transients for which ZTF is listed as the discovery data source on TNS to exclude cases where follow-up may have been motivated by external data, e.g. from another survey, which would bias the latency distribution.

Figure~\ref{fig:latency} shows the distribution of spectroscopic follow-up latencies for nearby SNe, highlighting those involving \BTSbot and that which was demonstrated by \BTSbotnearby for SN\,2024jlf. Latencies for SNe with human-triggered follow-up clearly cluster around $\Delta t_{\mathrm{spec}}=1$ day. This matches expectations for this time-period because visual inspection and triggering of high-priority follow-up typically only occur the morning after observations. When this is the case and the follow-up instrument is co-located with the survey (e.g. ZTF and SEDM), follow-up tends to occur $\sim$24 hours after discovery. This amount of latency or more is reflected in the results of numerous infant SN studies including those which deal with UV follow-up; see, e.g., Figure~2 of \cite{Irani+2024}.

This is not always the case, however. Scanners in Europe are able to easily scan in near real-time because the Palomar night coincides with morning and afternoon in Europe. This has resulted in countless instances of early follow-up triggered by European scanners yielding valuable results \citep[e.g.,][]{Yang+2021, Sollerman+2021}. In fact, much of the follow-up effort supporting \cite{Bruch+2021} and \citetalias{Bruch+2023} was enabled by scanners in Europe. Because the sample used here is bounded by 1 January 2021, it does not contain any SNe in \citetalias{Bruch+2023}, but the latency statistics are representative for human scanning between January 2021 and October 2024.

Comparing the \BTSbot latencies with that of human scanners in Figure~\ref{fig:latency} shows that most follow-up involving \BTSbot is much quicker: $\Delta t_{\mathrm{spec}}\ll1$ day. The event with the single lowest latency in this sample is SN\,2024jlf: $\Delta t_{\mathrm{spec}}\approx7$ minutes. Automating transient identification and the triggering of rapid response follow-up mitigates costly latency. With extremely small latencies observations can probe the mostly unexplored earliest phases, i.e. hours after first light, of extragalactic transients.

\section{Observations of SN 2024jlf} \label{sec:obs}

\subsection{Discovery and Classification} \label{sec:discovery}

SN\,2024jlf was first reported to TNS by \cite{24jlf_discovery_Hinds} at 10:29:59 on 28 May 2024 (modified Julian date; MJD 60458.44) using data from ZTF (internal name ZTF24aaozxhx), which measured its brightness to $g_{\textrm{ZTF}}=15.88\pm0.04$ mag. It has been localized in NGC\,5690 to \mbox{$\alpha_{\mathrm{J2000}}=14^{\mathrm{h}} 37^{\mathrm{m}} 42^{\mathrm{s}}_{^{\centerdot}}32$} \mbox{$\delta_{\mathrm{J2000}}=+02^{\degree}17' 04''_{^{\centerdot}}12$}. About three hours later, \cite{24jlf_classification_Hosseinzadeh} and the Global SN Project \citep{Howell_2017AAS} classified SN\,2024jlf as a young SN\,II and noted that it displayed weak flash features.

Before either TNS report, \BTSbot identified ZTF24aaozxhx as a genuine bright ($m_{\mathrm{peak}}<18.5$ mag) transient, saved it to an internal ZTF transient catalog, and triggered a high-priority follow-up request for photometry ($ugri$) and spectroscopy to SEDM at 07:53:40 (MJD 60458.33).\footnote{In the future, \texttt{BTSbot-nearby} discoveries will automatically be reported to TNS to facilitate community follow-up of these important targets.} SEDM began observing ZTF24aaozxhx just $\sim$7 minutes later at 08:00:51; this observation occurred only $+$0.7 days after first light (see Sec.~\ref{sec:lc_params} for details). Logs of Fritz activity, where ZTF partners scan new transient candidates, suggest that no astronomer in ZTF had viewed ZTF24aaozxhx before the spectroscopic observation had concluded. Following this discovery, we collected multi-wavelength, high-cadence follow-up observations with numerous facilities to probe the evolution of SN\,2024jlf.

\begin{table} 
    \begin{center} 
    \caption{Basic properties of SN\,2024jlf}
    \label{tab:basic_properties}
    \footnotesize
    \begin{tabularx}{0.46\textwidth}{c c}
        \toprule
        Property & Value \\
        \hline
        Distance modulus $\mu$ [mag] & $31.33\pm0.43$ \\
        Luminosity distance $D_L$ [Mpc] & $18.45\pm3.66$ \\
        $z_{\mathrm{SDSS}}$ & $0.00592\pm0.00002$ \\
        $\mathrm{EW}_{\mathrm{Na~ID}}$ [\r{A}] & $\NaIDEW\pm\NaIDEWunc$ \\
        $A_{V, \mathrm{host}}$ [mag] & $\AVhost\pm\AVhostunc$ \\
        $A_{V, \mathrm{MW}}$ [mag] & $\AVMW$ \\
        MJD of first light $t_{fl}$ & $60457.62\pm0.054$ \\
        $M_{u,\mathrm{peak}}$ [mag] & $-17.10^{+0.10}_{-0.10}$ \\
        $M_{g,\mathrm{peak}}$ [mag] & $-16.97^{+0.072}_{-0.071}$ \\
        $M_{r,\mathrm{peak}}$ [mag] & $-16.78^{+0.064}_{-0.064}$ \\
        $M_{i,\mathrm{peak}}$ [mag] & $-16.58^{+0.13}_{-0.14}$ \\
        Duration of flash features $\tau$ [d] & $1.3-1.8$\\
    \hline
    \end{tabularx}
    \end{center}
\end{table}

\subsection{Host Galaxy -- NGC 5690}
\label{sec:host}

SN\,2024jlf occurred in NGC\,5690, an edge-on spiral galaxy showing prominent dust lanes. \cite{deVaucouleurs+1991} classify NGC\,5690 as a possible Sc galaxy, with some uncertainty due to the edge-on orientation. The distance modulus to NGC\,5690 is measured to be $\mu=\DM\pm\DMunc$ mag by \cite{Sorce+2014} using the Tully-Fisher relation \citep{TullyFisher1977}; this corresponds to a luminosity distance of $D_L=\dist\pm\distunc$ Mpc. A Sloan Digital Sky Survey \citep[SDSS;][]{York+2000} spectrum of the galaxy's nucleus yields $z_{\textrm{SDSS}}=\z\pm\zunc$ \citep{Albareti+2017}. We adopt these values as the distance modulus, distance, and redshift to SN\,2024jlf, and they are summarized in Table~\ref{tab:basic_properties}.

NGC\,5690 has been imaged by the InfraRed Array Camera \citep[IRAC;][]{Fazio+2004} on the Spitzer Space Telescope (\textit{Spitzer}) through multiple \textit{Spitzer} programs: PID 14098 (PI O.\,Fox), PID 61066 (PI K.\,Sheth), and PID 10046 (PI D.\,Sanders). 
We stack the available 3.6 $\mu$m and 4.5 $\mu$m imagery to search for the progenitor of SN\,2024jlf but are unable to detect a source at the SN location.

\begin{figure}
    \begin{center}
    \includegraphics[width=1\columnwidth]{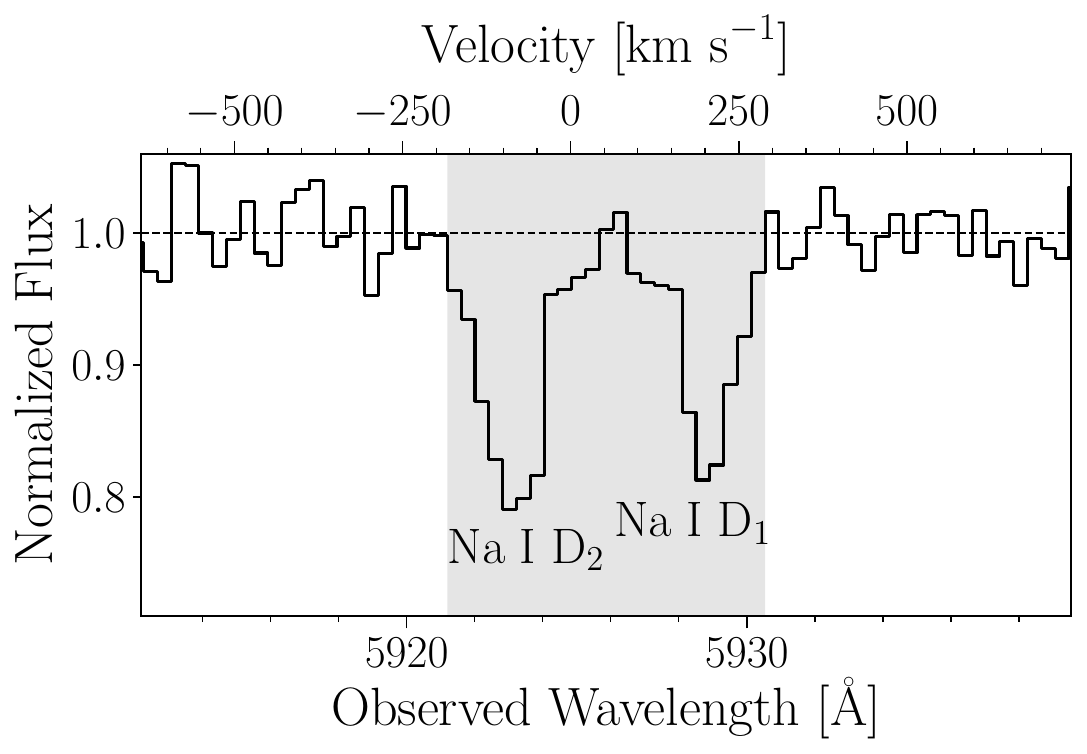}
    \caption{Na ID absorption doublet (shaded gray region) in a continuum-normalized spectrum of SN\,2024jlf (solid black line). The large equivalent width, measured to be $\NaIDEW\pm\NaIDEWunc$\;\r{A}, suggests significant extinction by the host galaxy. The velocity axis is defined relative to the \ion{Na}{1}\;D$_2$ line at the host redshift from SDSS. The sodium gas producing the absorption appears to move with $v=80$ km s$^{-1}$ relative to the host nucleus, where the SDSS fiber is placed.
    }
    \label{fig:NaID}
    \end{center}
\end{figure}

We use the equivalent width of the Na ID doublet as a proxy for the reddening toward the SN due to the host galaxy. This practice, although very common, has important caveats \citep{Poznanski+2011, Phillips+2013}. We directly integrate a continuum-normalized spectrum over the Na ID region (see Fig.~\ref{fig:NaID}) to compute an equivalent width of $\mathrm{EW}_{\mathrm{Na~ID}}=\NaIDEW\pm\NaIDEWunc$~\r{A}. The uncertainty reported is the standard deviation of 1000 trials of recomputing the equivalent width using Monte Carlo samples of the flux and its uncertainty. Using the relation
in \citealt{Stritzinger+2018}, we compute $A_{V,\mathrm{host}}=\AVhost\pm\AVhostunc$~mag. Section~\ref{sec:phot} describes how reddening correction is applied to the photometry given this $A_{V,\mathrm{host}}$. Figure~\ref{fig:NaID} also illustrates that the velocity of the sodium gas producing the absorption is moving at 80 km s$^{-1}$ relative to the galaxy nucleus, as measured by the SDSS spectrum.

\subsection{Photometric observations} \label{sec:phot}

\begin{figure*}[ht]
    \begin{center}
    \includegraphics[width=1\linewidth]{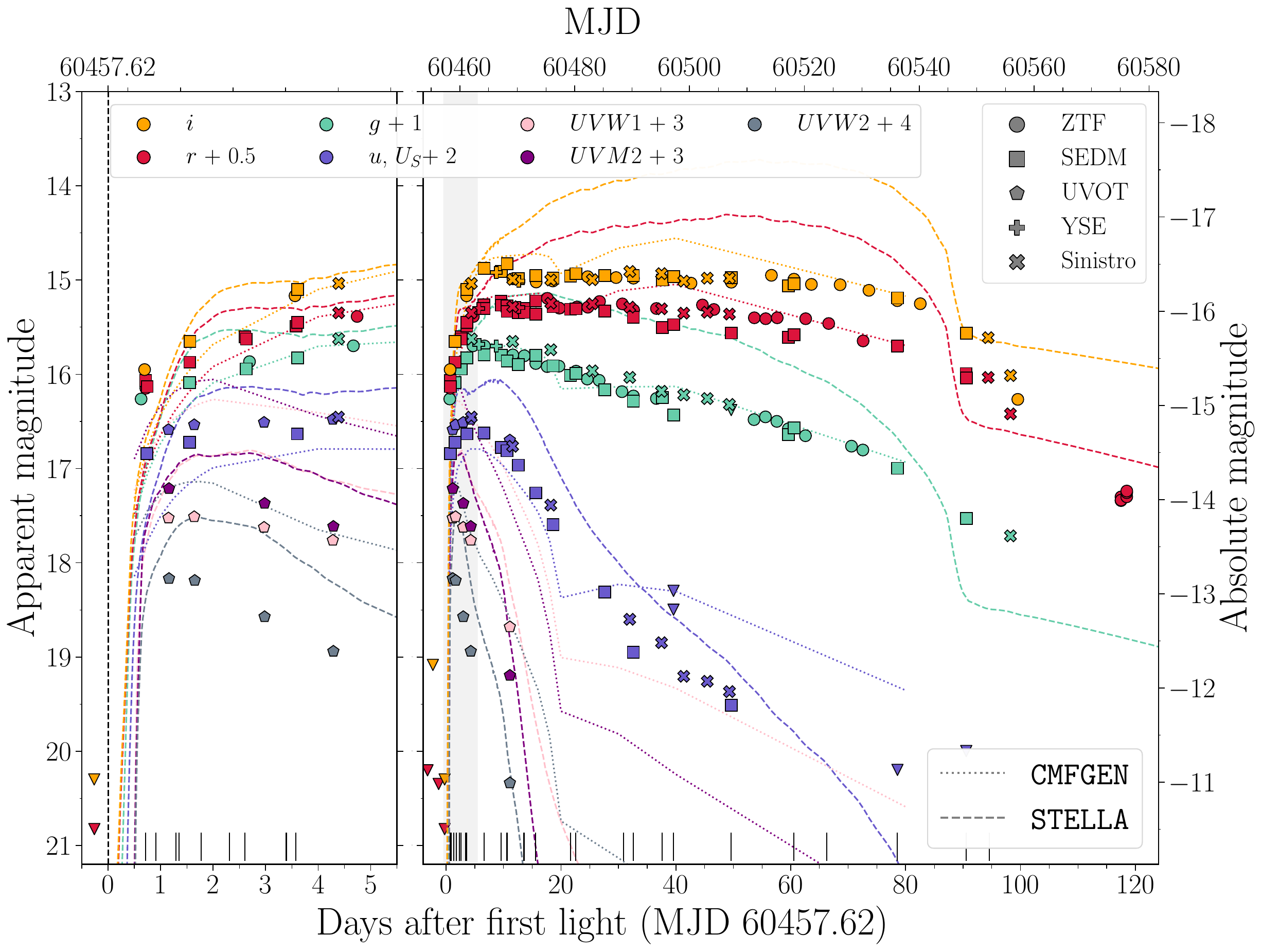}
    \caption{Multi-band light curve of SN\,2024jlf and best fit model light curves. \textit{Left panel} (and shaded region in \textit{right panel}): SN\,2024jlf rises exceptionally rapidly in all bands and begins to fade in the UV $\sim$24 hours after $t_{fl}$ (dashed black line). The \texttt{STELLA} model better reproduces the first five days after $t_{fl}$, but both models significantly underestimate the optical brightness $\sim$24 hours after $t_{fl}$. \textit{Right panel}: As a normal SN\,IIP, SN\,2024jlf maintains near peak-luminosity for $\sim$85 days before beginning to fade more rapidly. The \texttt{CMFGEN} model well reproduces the plateau luminosity across the optical bands, but the \texttt{STELLA} model strongly overestimates them.}
    \label{fig:lightcurve}
    \end{center}
\end{figure*}


The ZTF public and partnership surveys observed SN\,2024jlf from the initial discovery to when the field moved behind the Sun. We obtain forced point spread function (PSF) photometry from the ZTF forced photometry service \citep{Masci+2019, Masci+2023}, which we then process to calibrated flux measurements with the pipeline presented in A. A. Miller et al. (in prep.).

YSE \citep{Jones+2021} monitoring with the Pan-STARRS1 (PS1) telescope \citep{Kaiser+2002} produced deep non-detections shortly before the first detection by ZTF and several detections afterwards. All YSE data were processed using {\tt photpipe} \citep{Rest05} following methods described in \citet{Jones+2021}, including digital image subtraction with PS1 3$\pi$ templates \citep{Chambers16} using {\tt HOTPANTS} \citep{Becker15}.  All YSE photometry and pre-explosion limits presented here are derived from forced photometry in the difference images.
 
We conducted optical imaging of SN\,2024jlf with the SEDM Rainbow Camera (RC) onboard the P60 telescope \citep{Cenko+2006} and the Sinistro imager at Sliding Spring Observatory in the Las Cumbres Observatory (LCO) 1\,m telescope network \citep{Brown+2013}. SEDM RC observations were reduced with the \texttt{FPipe} automated subtraction pipeline \citep{Fremling+2016}. Sinistro images were reduced using {\tt photpipe} following similar methods to our YSE data.  As pre-explosion images were not available in each band, we did not perform image subtraction in the LCO imaging and instead report {\tt dophot} \citep{Schechter93} PSF photometry for all detections coincident with SN\,2024jlf in the unsubtracted images.  This potentially introduces a small systematic bias relative to the rest of our optical photometry, which is derived in difference imaging.

SN\,2024jlf was observed in the UV ($UVW1$, $UVW2$, $UVM2$) and near-UV ($U_S$) with the Ultraviolet/Optical Telescope \citep[UVOT;][]{Roming+2005} onboard the Neil Gerhels Swift Obervatory \citep[\textit{Swift};][]{Gehrels+2004}.  We processed all UVOT photometry of SN\,2024jlf using a {\tt Python}-based wrapper\footnote{\url{https://github.com/charliekilpatrick/Swift_host_subtraction}} for photometry tools in {\tt heasoft} v6.28 \citep{heasoft}. We optimally stacked each epoch and performed aperture photometry in each band using {\tt uvotsource}.

All photometry is corrected for Milky Way and host extinction using the \texttt{extinction} package \citep{Barbary_extinction}. We adopt the extinction law from \cite{Fitzpatrick+1999}, $R_{V,\mathrm{MW}}=R_{V,\mathrm{host}}=3.1$, and the dust map from \cite{Schlafly_Finkbeiner_2011} queried using the \texttt{dustmaps} package \citep{dustmaps_2018, dustmaps_2024}. The dust map yields $E(B-V)_{\mathrm{MW}}=0.039$ mag which corresponds to $A_{V,\mathrm{MW}}=\AVMW$ mag. The reddening from the host is applied using $A_{V,\mathrm{host}}=\AVhost\pm\AVhostunc$ mag, computed from the equivalent width of the Na ID doublet (see Sec.~\ref{sec:host}). Lastly, we inflate the magnitude uncertainties by 2\% to account for otherwise unconsidered systematics. All photometry is reported in Table~\ref{tab:phot}.

\begin{deluxetable}{ccccc} \label{tab:phot}
    \tabletypesize{\scriptsize}
    \tablewidth{0pt}
    \tablecaption{Optical and UV Photometry of SN\,2024jlf.}
    \tablehead{
        \colhead{$t_\mathrm{obs}$} &
        \colhead{Filter} &
        \colhead{$m$} &
        \colhead{$\sigma_m$} &
        \colhead{Telescope/} \\
        \colhead{$[\textrm{MJD}]$} &
        \colhead{} &
        \colhead{$[\textrm{mag}]$} &
        \colhead{$[\textrm{mag}]$} &
        \colhead{Instrument}
    }
    \startdata
    60458.25 & $g_\mathrm{ZTF}$ & 15.878 & 0.0133 & P48/ZTF \\
    60458.32 & $i_\mathrm{ZTF}$ & 16.239 & 0.0169 & P48/ZTF \\
    60458.33 & $r'$ & 15.991 & 0.0405 & P60/SEDM \\
    60458.33 & $r'$ & 15.991 & 0.054 & P60/SEDM \\
    60458.34 & $r_\mathrm{ZTF}$ & 16.055 & 0.0236 & P48/ZTF \\
    \enddata
    \tablecomments{Observed magnitudes in the ZTF, YSE, UVOT, SEDM, and Sinistro passbands. Correction for Galactic extinction has not been applied.\\(This table is available in its entirety in machine readable form.)}
\end{deluxetable}

The final light curve is shown in Figure~\ref{fig:lightcurve}. The right panel shows the full light curve including a steep rise and $\sim$85 day plateau. The left panel zooms in on the early light curve and makes clear the exceptionally quick rise of SN\,2024jlf. Short vertical lines across the bottom of the figure represent each epoch of spectroscopic observations. Best-fit model light curves are also shown and discussed in Sections~\ref{sec:cmfgen_phot} and \ref{sec:moriya_models}.

\begin{figure}
    \begin{center}
    \includegraphics[width=1\columnwidth]{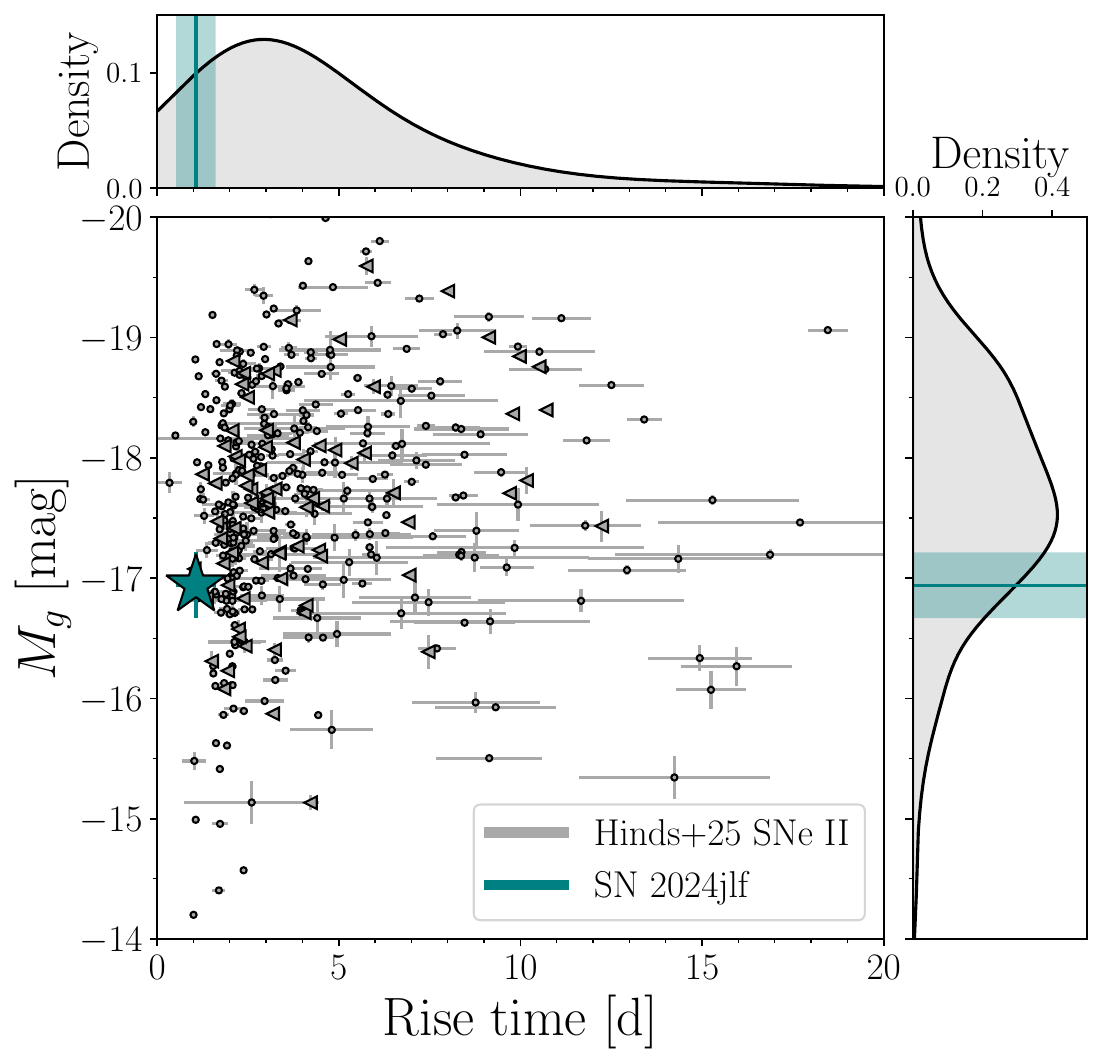}
    \caption{Comparison of SN\,2024jlf (teal) $g$-band light curve properties with those of a SN\,II sample (gray; Hinds et al., in prep.). SNe with poor light curve coverage (see Hinds et al. for definition) only have upper limits on rise time, which are shown as arrowheads. SN\,2024jlf rises quicker than $\sim$90\% of the sample despite peaking slightly fainter than average. Accelerated rise times are often seen in SNe\,II with CSM interaction.}
    \label{fig:rise}
    \end{center}
\end{figure}

We compare the observed light curve properties with that of 459 ZTF SNe\,II from Hinds et al. (in prep.). Figure~\ref{fig:rise} shows the $g$-band absolute magnitude and rise time (defined as time from 25\% to 75\% of maximum flux), both inferred from Gaussian process fits performed in Hinds et al.\ (in prep.). We find that SN\,2024jlf rises quicker than $\sim$90\% of SNe\,II in the sample and peaks in the $g$-band slightly fainter than average. This extremely rapid rise is likely a product of early-time flux excess originating from CSM interaction. 

This is consistent with the findings of \citetalias{Jacobson-Galan+2024_FMII}, where it is shown that SNe\,II with weak/intermediate CSM interaction tend to rise quicker than those without.

\subsubsection{Basic light curve parameters}
\label{sec:lc_params}

We fit low order polynomials to each of the optical bands to infer their peak magnitudes. The values reported in Table~\ref{tab:basic_properties} are the median and one sigma bounds from performing 10,000 Monte Carlo samples of the polynomial parameter uncertainties produced by the fit. Although each instrument uses slightly different filter systems, i.e. $g_{\textrm{ZTF}}$ for ZTF, $g_{\textrm{PS1}}$ for YSE, and $g'$ for SEDM and Sinistro, we approximate these as the same when performing this modeling. 

\begin{figure}
    \begin{center}
    \includegraphics[width=1\columnwidth]{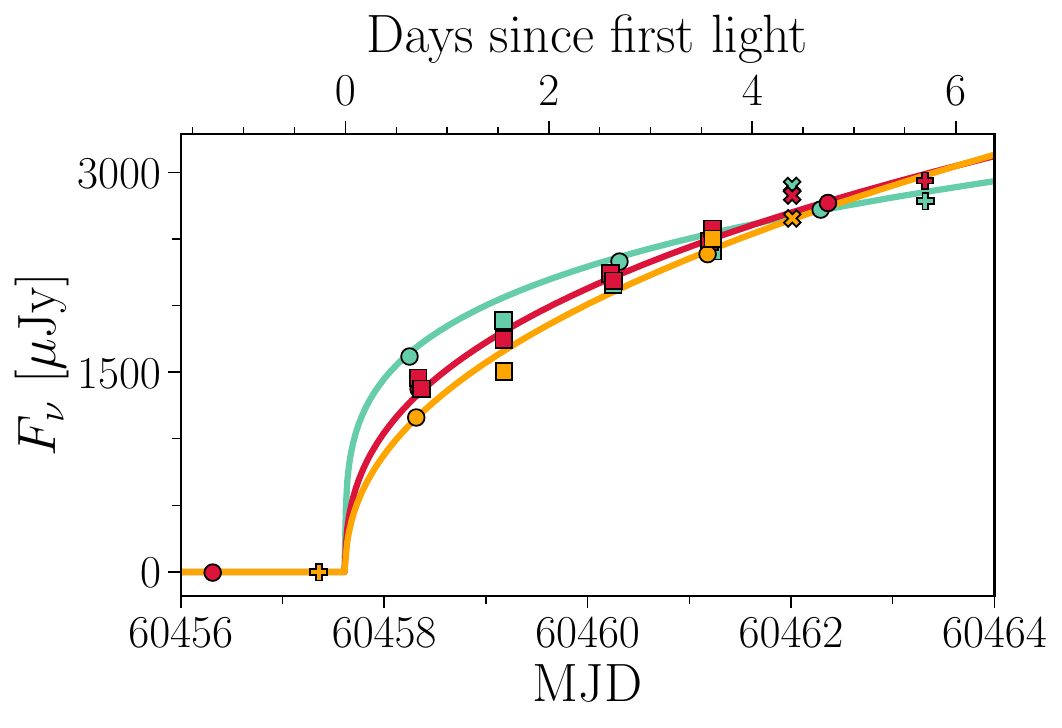}
    \caption{Joint power-law fits to the early $gri$ photometry of SN\,2024jlf (same colors and markers as Fig.~\ref{fig:lightcurve}) produce a time of first light estimate $t_{fl}=60457.62\pm0.054$. Non-detections shortly before first light enable a precise estimation.}
    \label{fig:t_fl}
    \end{center}
\end{figure}

We also jointly fit power laws to the early $gri$ photometry to infer the time of first light. The power laws are each in the form 
\begin{equation}
    F_{\nu}(t)=
    \begin{cases}
        a(t-t_{fl})^b & \text{if } t \geq t_{fl}\\
        0 & \text{if } t < t_{fl}
    \end{cases}
\end{equation}
where $a$ and $b$ can vary for each filter, but $t_{fl}$ is fixed across all filters \citep{Miller+2020_earlyIa}.
Figure~\ref{fig:t_fl} shows the power laws which result from this procedure. The very late YSE non-detection and the rapid follow-up allow us to constrain the time of first light to $t_{fl}=60457.62\pm0.054$\,MJD. We adopt this value throughout this study.

\subsection{Spectroscopic observations} \label{sec:spec}

\begin{figure}
    \begin{center}
    \includegraphics[width=0.92\columnwidth]{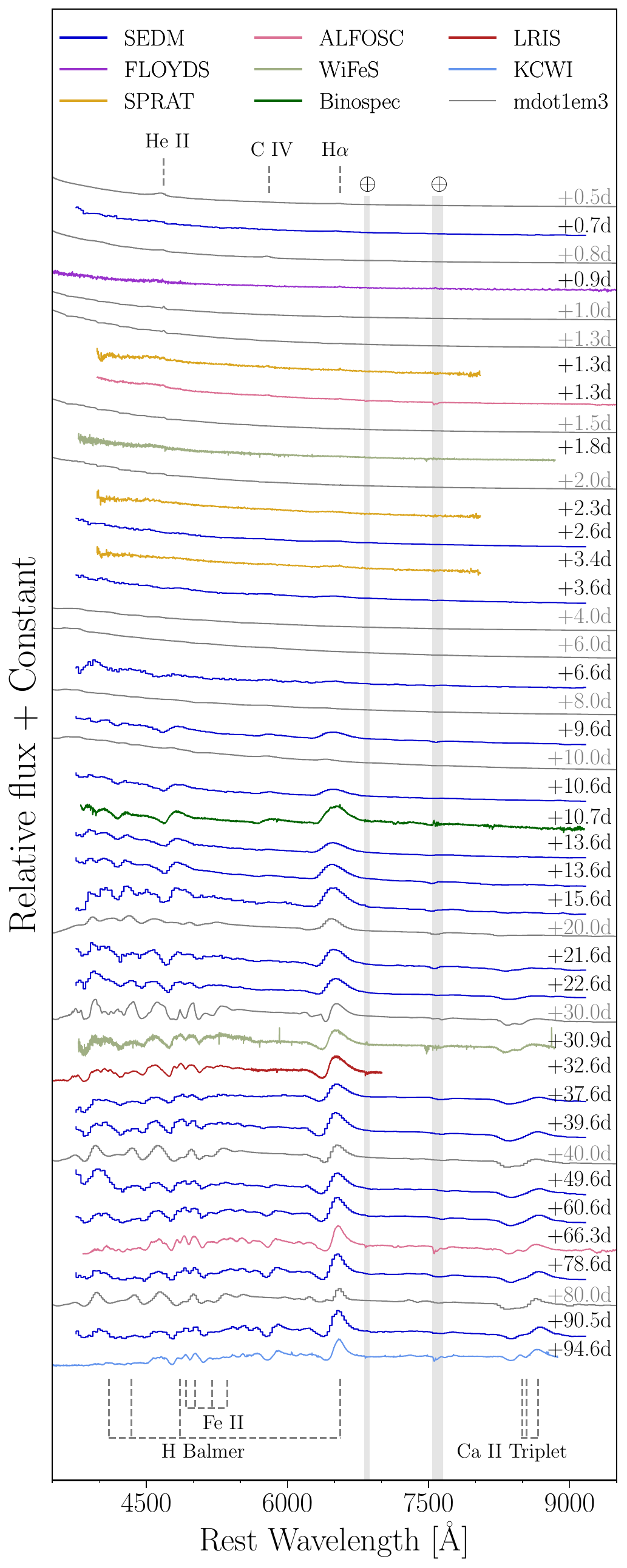}
    \caption{Full spectral series of SN\,2024jlf (colored lines) and best matched model spectra (gray lines). SN\,2024jlf is a normal SN~IIP showing flash features starting from a spectrum acquired just $+0.7$ days after $t_{fl}$ (see Fig.~\ref{fig:flash_features}). Over the next $\sim$100 days, the SN develops features typical of SNe~IIP, like broad Balmer series features.}
    \label{fig:spec_series}
    \end{center}
\end{figure}

\begin{table}
    \centering
    \caption{Spectroscopic observations of SN\,2024jlf}
    \label{tab:spectra}
    \footnotesize
    \begin{tabularx}{0.46\textwidth}{c c c c c}
        \toprule
         $t_{\mathrm{obs}}$ & Phase & Telescope/ & R & $\lambda$ Range \\
         $[\textrm{MJD}]$ & $[\textrm{d}]$ & Instrument & ($\lambda/\Delta\lambda$) & \r{A} \\
        \hline
         60458.33 & 0.7 & P60/SEDM &     100 & 3770--9220 \\
         60458.53 & 0.9 & FTN/FLOYDS-N & 550 & 3350--10000 \\
         60458.91 & 1.3 & LT/SPRAT &     350 & 4000--8100 \\
         60458.97 & 1.3 & NOT/ALFOSC &   360 & 4000--9680 \\
         60459.39 & 1.8 & ANU/WiFeS &   3000 & 3200--9565 \\
         60459.93 & 2.3 & LT/SPRAT &     350 & 4000--8100 \\
         60460.23 & 2.6 & P60/SEDM &     100 & 3770--9220 \\
         60461.01 & 3.4 & LT/SPRAT &     350 & 4000--8100 \\
         60461.20 & 3.6 & P60/SEDM &     100 & 3770--9220 \\
         60464.20 & 6.6 & P60/SEDM &     100 & 3770--9220 \\
         60467.18 & 9.6 & P60/SEDM &     100 & 3770--9220 \\
         60468.22 & 10.6 & P60/SEDM &     100 & 3770--9220 \\
         60468.27 & 10.7 & MMT/Binospec & 1340 & 3820--9210 \\
         60471.18 & 13.6 & P60/SEDM &     100 & 3770--9220 \\
         60471.21 & 13.6 & P60/SEDM &     100 & 3770--9220 \\
         60473.20 & 15.6 & P60/SEDM &     100 & 3770--9220 \\
         60479.26 & 21.6 & P60/SEDM &     100 & 3770--9220 \\
         60480.21 & 22.6 & P60/SEDM &     100 & 3770--9220 \\
         60488.50 & 30.7 & ANU/WiFeS &   3000 & 3200--9565 \\
         60490.24 & 32.6 & Keck/LRIS &    1400 & 3100--5730 \\
         60490.24 & 32.6 & Keck/LRIS &    8500 & 5400--7050 \\
         60495.23 & 37.6 & P60/SEDM &     100 & 3770--9220 \\
         60497.18 & 39.6 & P60/SEDM &     100 & 3770--9220 \\
         60507.23 & 49.6 & P60/SEDM &     100 & 3770--9220 \\
         60518.17 & 60.6 & P60/SEDM &     100 & 3770--9220 \\
         60523.93 & 66.3 & NOT/ALFOSC &   360 & 4000--9680 \\
         60536.17 & 78.6 & P60/SEDM &     100 & 3770--9220 \\
         60548.15 & 90.5 & P60/SEDM &     100 & 3770--9220 \\
         60552.23 & 94.6 & Keck/KCWI &    900 & 3275--8925 \\
    \hline
    \end{tabularx}
\end{table}

We conducted a thorough spectroscopic follow-up campaign to monitor the evolution of SN\,2024jlf out to $+$94.6 days after $t_{fl}$. Spectroscopic follow-up was conducted with numerous facilities: SEDM \citep{Blagorodnova+2018, Kim+2022} on the P60 telescope \citep{Cenko+2006}; the Spectrograph for the Rapid Acquisition of Transients \citep[SPRAT;][]{Piascik+2014} on the robotic 2\,m Liverpool Telescope \citep[LT;][]{Steele+2004}; the Alhambra Faint Object Spectrograph and Camera (ALFOSC)\footnote{\url{https://www.not.iac.es/instruments/alfosc/}} on the 2.56\,m Nordic Optical Telescope (NOT); Binospec \citep{Fabricant+2019} on the 6.5\,m MMT telescope; the Low Resolution Imaging Spectrometer \citep[LRIS;][]{Oke+1995} on the W. M. Keck Observatory's Keck I 10\,m telescope; the Keck Cosmic Web Imager \citep[KCWI;][]{Martin+2010, Morrissey+2018} on the Keck II 10\,m telescope; and the Wide-Field Spectrograph \citep[WiFeS;][]{Dopita07,Dopita10} on the Australian National University 2.3\,m (ANU) Advanced Technology Telescope (ATT) at Siding Spring Observatory. We also include the FLOYDS-N spectrum uploaded to TNS by \cite{24jlf_classification_Hosseinzadeh}. 

The SEDM observations are reduced by the custom \texttt{pySEDM} package \citep{Rigault+2019}. Spectra from other facilities are reduced with standard procedures (see \citealt{Matheson+2000}). Reduction of MMT/Binospec and Keck/LRIS spectra uses \texttt{pypeit} \citep{pypeit}; LT/SPRAT reduction uses the custom FRODOSpec pipeline \citep{Barnsley+2012}; NOT/ALFOSC reduction uses a custom \texttt{pypeit}-based script;\footnote{\url{https://gitlab.com/steveschulze/pypeit_alfosc_env}} Keck/KCWI reduction uses the official data reduction pipeline;\footnote{\url{https://kcwi-drp.readthedocs.io/}} and ANU/WiFeS reduction uses {\tt PyWiFeS} \citep[see][]{Childress14,Carr24}.\footnote{\url{https://github.com/PyWiFeS/pipeline}} Our spectroscopic observations are summarized in Table~\ref{tab:spectra}.

Figure~\ref{fig:spec_series} shows our full spectral series of SN\,2024jlf.
We find that SN\,2024jlf exhibits features and evolution typical of SNe\,IIP: a mostly featureless blue continuum followed by prominent P-Cygni lines of the hydrogen Balmer series and later of He. A broad H$\alpha$ profile first emerges in the $+2.3$ day SPRAT spectrum. H$\beta$ and bluer Balmer series lines become visible shortly afterwards; the Binospec spectrum at $+10.7$ days shows P-Cygni profiles in H$\alpha$, H$\beta$, H$\gamma$, H$\delta$, and possibly H$\epsilon$. He I $\lambda5876$ is also visible in a P-Cygni profile starting in the $+9.6$ day spectrum. Fe~II absorption and the Ca near-infrared triplet are visible starting around three weeks after first light. Model spectra, shown as gray lines, are discussed in Section~\ref{sec:cmfgen_spec}.

\begin{figure}
    \begin{center}
    \includegraphics[width=1\columnwidth]{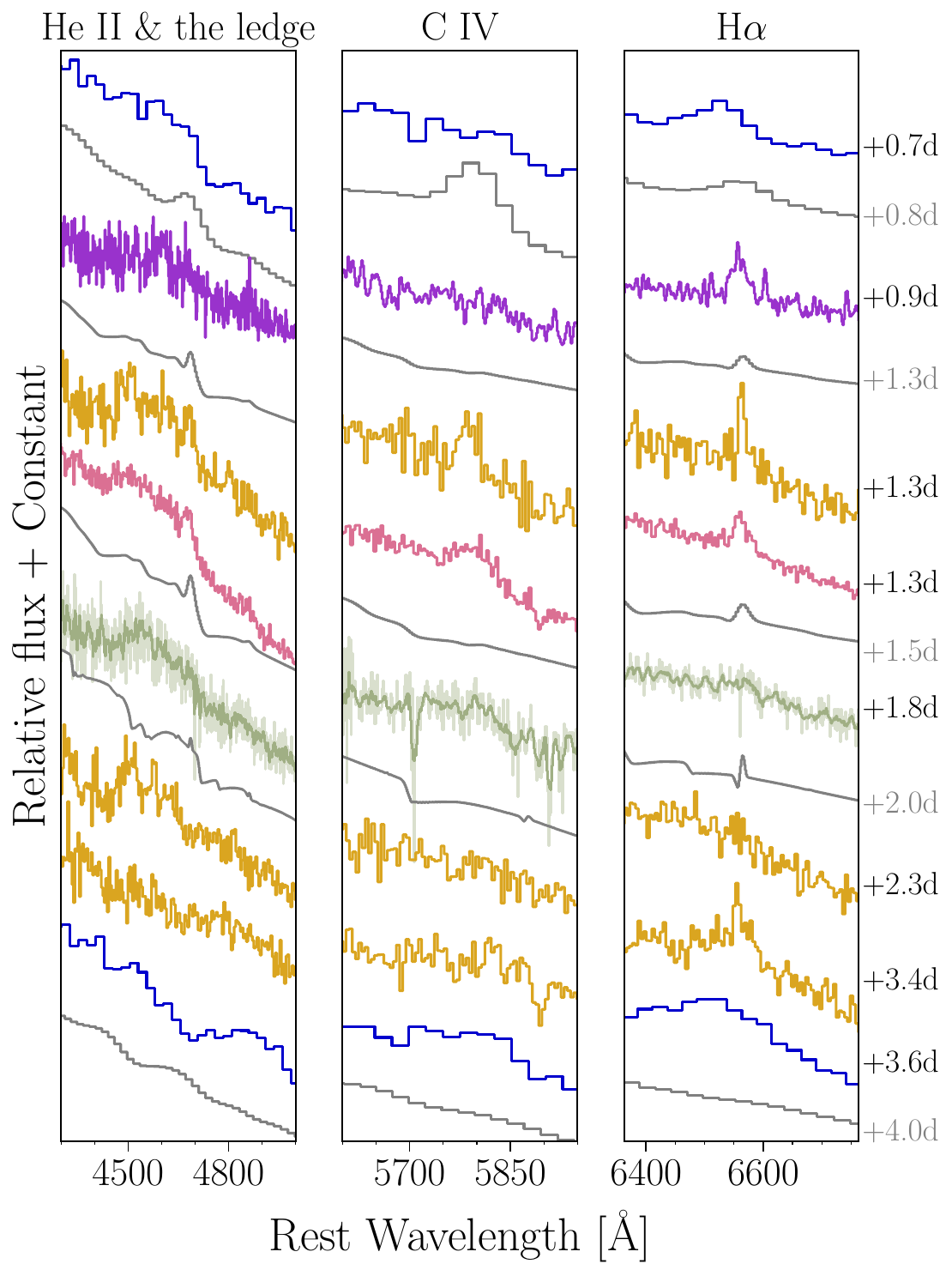}
    \caption{Early spectra of SN\,2024jlf (colored lines) and best matched model (gray lines) around regions showing flash features. \ion{He}{2}, \ion{C}{4}, H$\alpha$, and possibly H$\beta$ can be seen in narrow, short-lived emission. The WiFeS spectrum is shown un-smoothed (light green) and smoothed with a moving average filter (dark green) because it has greater noise around the important \ion{He}{2} feature. All flash features, including \ion{He}{2}, are present in the $+1.3$ day spectra but absent in the $+1.8$ day spectrum: $1.3<\tau\textrm{ [d]} < 1.8$.}
    \label{fig:flash_features}
    \end{center}
\end{figure}

Figure~\ref{fig:flash_features} shows the regions around main spectral lines in the early spectra of SN\,2024jlf which exhibit features characteristic of SNe~II undergoing CSM interaction: \ion{He}{2} $\lambda$4686 and the ledge, \ion{C}{4} $\lambda\lambda$5801, 5812, and H$\alpha$. We also possibly detect narrow H$\beta$ emission in the $+0.9$ d FLOYDS spectrum. \ion{He}{2} is very common in flashing SNe and its presence is often used to define a SN as flashing (e.g., \citealt{Khazov+2016, Bruch+2021}; \citetalias{Bruch+2023}). The ledge-shaped feature around 4450--4700 \r{A}, visible in the first six epochs of spectroscopy, is typically interpreted as a sign of CSM interaction and has been seen in many interacting SNe including SN\,2013fs \citep{Yaron+2017} and others in the samples of \citetalias{Bruch+2023} and \citetalias{Jacobson-Galan+2024_FMII}. Its origin is not precisely known although it has been attributed to (i) a blend of high-ionization lines including those of carbon, nitrogen, and oxygen \citep[e.g.,][]{Soumagnac+2020, Bruch+2021, Pearson+2023}; (ii) high-velocity H$\beta$ \citep[e.g.,][]{Pastorello+2006}; and (iii) blueshifted \ion{He}{2} $\lambda$4686 \citep[e.g.,][]{Dessart+2017, Bullivant+2018, Chugai+2023}. See \cite{Pearson+2023} for further discussion and references on the ledge feature.

\begin{figure}
    \begin{center}
    \includegraphics[width=1\columnwidth]{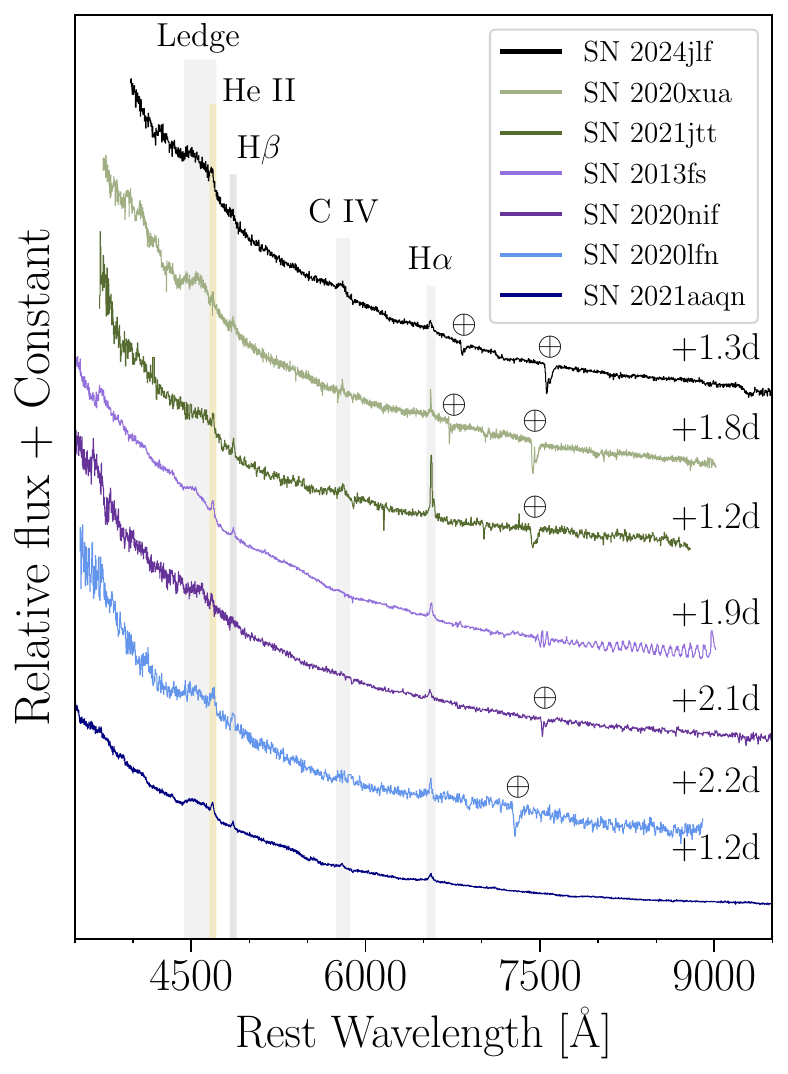}
    \caption{Comparing SN\,2024jlf and spectroscopically similar SNe from the \citetalias{Jacobson-Galan+2024_FMII} class 3 sample. All show narrow emission in \ion{He}{2}, H$\alpha$, and some additional emission feature(s) in the region of the ledge. Most events also have narrow \ion{C}{4} and/or H$\beta$ emission.}
    \label{fig:spec_comparison}
    \end{center}
\end{figure}

Figure~\ref{fig:spec_comparison} shows that SN\,2024jlf is also spectroscopically very similar to a number of previous events, selected from the class 3 sample in \citetalias{Jacobson-Galan+2024_FMII}. These other SNe (SN\,2020xua, \citealt{Terreran+2020_2020xua}; SN\,2021jtt, \citealt{Angus+2021_2021jtt}; SN\,2013fs, \citealt{Yaron+2017}; SN\,2020nif, \citealt{Hiramatsu+2020_2020nif}; SN\,2020lfn, \citealt{Izzo+2020_2020lfn}; and SN\,2021aaqn, \citealt{Taggart+2021_2021aaqn}) all show narrow \ion{He}{2}, H$\alpha$, and a variety of feature shapes in the region of the ledge. Most also show \ion{C}{4} emission (SNe\,2020xua, 2021jtt, 2021aaqn), and, although it is not obvious in SN\,2024jlf, nearly all others show narrow H$\beta$ emission as well (SNe\,2020xua, 2021jtt, 2013fs, 2020lfn, and 2021aaqn).

The duration of flash/IIn-like features is key to constraining the extent of the dense CSM.
\citetalias{Bruch+2023} define the flash timescale $\tau$ as the duration for which the \ion{He}{2} $\lambda$4686 feature persists. The median $\tau$ in their golden flasher sample is $5.4\pm2.7$ days, including some events with $\tau\lesssim2$ days. In all panels of Figure~\ref{fig:flash_features}, we can clearly identify the narrow features in the $+1.3$ day spectra from SPRAT and ALFOSC, and even earlier in the case of H$\alpha$, but they all disappear by the $+1.8$ day WiFeS spectrum. We thus find $1.3<\tau~\mathrm{[d]}<1.8$. There appears to be a narrow H$\alpha$ feature in the $+3.4$ day SPRAT spectrum. In Appendix~\ref{app:flash_2d} (Figure~\ref{fig:flash_2d}), we confirm from the reduced 2D spectra that this is contamination from the host by modeling the SN and galaxy continua and inspecting the flux residuals. This contamination is not present in all SPRAT spectra because its presence and strength is dependent on the observing conditions and slit position of each observation. 

\citetalias{Jacobson-Galan+2024_FMII} define $t_{\mathrm{IIn}}$ as the duration for which the narrow features show symmetric Lorentzian wings. This terminates once broad absorption profiles develop in the Balmer features, indicating that the CSM optical depth has decreased enough to allow the fastest moving SN ejecta to emerge. The parameter $t_{\mathrm{IIn}}$ is valuable because it is sensitive to the CSM density profile. The quality of our spectra and the strength of the flash features in SN\,2024jlf do not, however, allow us to unambiguously identify when the Lorentzian wings subside. When this is the case, \citetalias{Jacobson-Galan+2024_FMII} scale the $t_{\mathrm{IIn}}$ of a very similar SN with well-constrained $t_{\mathrm{IIn}}$ (e.g., SN\,2013fs) by the ratio of the epochs of the similar spectra (see \citetalias{Jacobson-Galan+2024_FMII} Sec.~3.2 for more details). This extrapolates the evolution of an event with very well constrained $t_{\mathrm{IIn}}$ and assumes it is proportionally consistent with that of the given SN. Following this procedure with our $+$1.3 day ALFOSC and the $+$1.9~day SN\,2013fs spectrum (see Fig.~\ref{fig:spec_comparison}), we infer $t_{\mathrm{IIn}}=0.96\pm0.34$\,d for SN\,2024jlf. Although this indirect inference is imperfect, it nonetheless appears to find a value consistent with the observations.

\section{Matching with \texttt{CMFGEN} models} \label{sec:cmfgen_models}

We match our spectral series and UV/optical light curve of SN\,2024jlf to models produced by \cite{Dessart+2017} and \cite{Dessart_WJG_2023}. Radiative hydrodynamics for these models were performed with the \texttt{HERACLES} code \citep{Gonzalez+2007}, and post-processing on specific \texttt{HERACLES} snapshots was performed with the radiative transfer code \texttt{CMFGEN} \citep{Hillier_Dessart_2012}.

The matching process involves comparing properties of the model and observed spectra: (i) which species are seen in narrow emission and their relative strengths; (ii) the duration of the flash features $\tau$; and (iii) the relative phases of when typical SN\,IIP features (e.g., broad H$\alpha$) develop.
Because there are relatively few models ($<$20) and they are quite heterogeneous across these three properties of interest, a quantitative search for the best fit model is unnecessary. By visual inspection, we find that the \texttt{mdot1em3} model best matches the spectral properties of SN\,2024jlf; its physical properties are discussed in Section~\ref{sec:discussion}.

\subsection{Comparison to model spectra}
\label{sec:cmfgen_spec}

We degrade the model spectra to facilitate comparison with observed spectra of varying spectral resolutions. This is done by convolving the model spectrum with a Gaussian kernel corresponding to the spectral resolution of the observed spectrum nearest in phase to the model spectrum. This process, and its effect on the visibility of narrow features in the ultra-low resolution SEDM spectra ($R\sim100$), is discussed in depth in \citetalias{Bruch+2023}. We also scale the continuum of the early (phase $<10$\,d) model spectra to match that of the nearest observed spectrum; this is not necessary for any of the later spectra.

Figure~\ref{fig:flash_features} compares the early spectral series of SN\,2024jlf (colored lines) with that of \texttt{mdot1em3} (gray lines). The model spectra show flash features in \ion{He}{2}, \ion{C}{4}, and H$\alpha$, as do the observed spectra, and no additional flash features appear in the model spectra. We find $2<\tau~\mathrm{[d]}<4$ for \texttt{mdot1em3}, which is nearly consistent with our constraints on $\tau$ for SN\,2024jlf: $1.3<\tau~\mathrm{[d]}<1.8$. The primary deviation between the observations and the model is that the model only develops a broad H$\alpha$ feature after $\sim$10 days, while we observe a similar feature emerging in the $+2.3$ day spectrum of SN\,2024jlf (see Fig.~\ref{fig:spec_series}).

There are minor deviations between the  model spectra and the observations: (i) The $+0.8$ day model spectrum shows a narrow \ion{He}{2} feature and no ledge while the $+0.7$ day SEDM spectrum shows a ledge without obvious narrow \ion{He}{2}. (ii) The model's \ion{C}{4} feature is prominent starting from $+0.8$ days despite not being visible until $+1.3$ days in SN\,2024jlf. (iii) The $+0.7$ day SEDM spectrum shows blue-shifted H$\alpha$ while the model shows no such shift. SN\,2023ixf also showed a blue-shifted H$\alpha$ profile in early high resolution spectroscopy \citep{Dickinson+2024}, although the shift only developed days after $t_{fl}$. Higher resolution spectroscopy would have been necessary to further investigate this feature. Despite these differences, all of the key features (\ion{He}{2}, \ion{C}{4}, and H$\alpha$) disappear on similar timescales in the model and observations, and some of the discrepancies could be explained by insufficient signal in the observations.

Across our full spectral series (see Fig.~\ref{fig:spec_series}), no other major deviations are apparent. The model well reproduces the Balmer series features and, once they appear, the \ion{Fe}{2} and \ion{Ca}{2} lines.

\subsection{Comparison to model photometry}
\label{sec:cmfgen_phot}

The corresponding model light curve (dotted lines) is compared to that of SN\,2024jlf (points) in Figure~\ref{fig:lightcurve}. The left panel makes clear that \texttt{mdot1em3} rises far slower in the optical and near-UV ($griuU_S$) than SN\,2024jlf. The optical bands are all underestimated through the first $7$ days after $t_{fl}$. In the UV bands ($UVW1$, $UVM2$, $UVW2$), however, this pattern is reversed; \texttt{mdot1em3} initially rises quicker than SN\,2024jlf and uniformly overestimates the UV flux.
After the initial rapid UV rise, \texttt{mdot1em3} then peaks later in the (near-)UV bands than SN\,2024jlf.

The right panel shows that peak luminosity is achieved at a very similar phase in the optical bands. The $r$- and $i$-bands are exceptionally well fit beyond $\sim$10 days after $t_{fl}$, but the $g$-band light curve of \texttt{mdot1em3} slightly overestimates the SN flux by $\sim$0.2 mag across the plateau. As was apparent in the early light curve, the UV flux from SN\,2024jlf is strongly overestimated by the model. Although the initial UV rise in \texttt{mdot1em3} is quicker than observed, the decline rate of the model is slower than observed.

The aspects of disagreement between the model and observations cannot be reconciled by changing the amount of host attenuation we correct for. Decreasing the $A_{V,\mathrm{ host}}$ from 0.636 to 0.4 causes the $gri$-bands to fit well from three days after $t_{fl}$ onward, but the early rise rate is still poorly reproduced and the overestimations in the near-UV and UV are significantly worsened. Increasing $A_{V,\mathrm{ host}}$ to 0.8 makes the $UVM2$ and $UVW2$ bands fit well between two and five days after $t_{fl}$, but (i) $UVW1$ is still overestimated; (ii) the fit to the early optical rise is worsened; and (iii) the overestimates in the near-UV and $g$-bands are greater.




\section{Matching with \texttt{STELLA} models} \label{sec:moriya_models}

We also compare the observations with a grid of $\sim$228,000 SN~II models from \cite{Moriya+2023}. These models are produced with the one-dimensional radiation hydrodynamics code \texttt{STELLA} \citep{Blinnikov+1998, Blinnikov+2000, Blinnikov+2006} and adopt RSG progenitor models from \cite{Sukhbold+2016}.
Although the models do produce SEDs well sampled in time, the SEDs do not have sufficient wavelength resolution to compare against observed spectra. For this reason, only light curves are available to compare against observations.

The models in the \cite{Moriya+2023} grid exhibit an artifact in the early light curve delaying first light. This artifact arises from the initial condition of the numerical simulation wherein
photons diffuse out of the CSM before the shock breakout. The ``zero" phase in these model light curves is the time at which thermal energy is injected into the simulation, but the actual shock breakout happens roughly 1--3 days later. \citet{Moriya_Singh_2024} handle this by recalculating the goodness-of-fit metric for each model by iterating through a grid of small adjustments to the provided phases.
Instead, we compute offsets between the provided zero epoch and the time of first light $t_{fl}$ by identifying when the light curve begins its primary brightening phase. 
These offsets are applied to the phases of the model's photometry, so that we can fix $t_{fl}$ before computing goodness-of-fit metrics.
We evaluate the fit of each model to the observed $g$-band light curve by computing the $\chi^2$ statistic, which quantifies the agreement between the model light curve and the observed data. This model-fitting approach quantifies how well each model reproduces the observed light curve while accounting for observational uncertainties. Models are ranked based on their $\chi^2$ values, with a lower $\chi^2$ indicating a better agreement. This approach enables a quantitative ranking of the $\sim$228,000 models in the grid. The physical properties of the best fit model are discussed in Section~\ref{sec:discussion}.\footnote{The best fit model is named \texttt{s10ni0p04hm6\_m4.0b5.0r8e14e15}.}


The best fit model does not capture SN\,2024jlf's exceptionally rapid optical and near-UV rise in the first $\sim$24 hours after $t_{fl}$. From $2-7$ days, however, the optical and near-UV light curves are quite well matched. The UV flux is underestimated in the first epoch of \textit{Swift}/UVOT photometry, but the decline rate is well matched by the final epoch. 

The optical flux during the plateau, $\sim$7 days after $t_{fl}$ and onward, is overestimated. The brighter plateau light curves from the \citet{Moriya+2023} grid can be attributed to the progenitor models derived by \citet{Sukhbold+2016}, combined with the inherent degeneracies in light curve modeling \citep{Goldberg+2019}. The progenitors are evolved employing a low mixing length parameter, leading to systematically larger progenitor radii for a given ZAMS mass, which enhances the recombination radius and increases the plateau brightness \citep{Moriya_Singh_2024}. In addition, $\rm^{56}$Ni being mixed to half of the H-envelope in the progenitor models of \citet{Sukhbold+2016} leads to a brighter plateau. This is because $\rm^{56}$Ni starts diffusing out much earlier than the end of the plateau phase, as was seen in SN~2016gfy \citep{Singh+2019a}. The combination of these effects causes the model's optical light curves, particularly in the redder bands, to peak later than observed. In fact, the modeled peaks occur after the observed peaks across all bands. Although other models in the \cite{Moriya+2023} grid better match the flux on the plateau, we prefer this model as it fits the rise better than any others. The CSM interaction most significantly influences the rise and the UV light curves, so this model should provide the best estimate of the CSM properties.

Since the models from \cite{Moriya+2023} extend past the end of the plateau, we can assess their fit of the plateau duration and the $^{56}$Co tail. The best fit model's plateau is $\sim$10 days shorter than that of SN\,2024jlf, and the model's $r$-band tail flux is within $\sim$0.2 mag of the ZTF photometry $+120$ days after $t_{fl}$.

We modulate the amount of host attenuation as an attempt to improve the match. Increasing $A_{V,\mathrm{ host}}$ to 0.9 causes the near-UV and $g$-band flux to match the model very well beyond $\sim$10 days after $t_{fl}$, but doing so (i) ruins the previously good fit to the optical data between 24 and 72 hours after $t_{fl}$; (ii) causes the UV to be systematically and significantly underestimated; and (iii) does not alleviate the poor fit to the $ri$-band plateau flux. Decreasing $A_{V,\mathrm{ host}}$ worsens nearly all aspects of the match. We conclude that alternative values of $A_{V,\mathrm{ host}}$ do not provide a significantly better match between this model and our observations.

\section{Discussion} \label{sec:discussion}

We have matched observations of SN\,2024jlf to models from grids produced by two different radiative hydrodynamics codes. Here, we compare the ability of the best matched models from these very different grids to reproduce the observed properties of SN\,2024jlf and compare the physical properties inferred from the models.

\subsection{Comparison of model fits}
\label{sec:model_comp}

We compare only the models' ability to reproduce the photometry and not the spectroscopy because the \texttt{STELLA} SEDs are not of sufficient wavelength resolution to compare against observations.


Neither model better reproduces the entire optical light curve than the other model. The best matched \texttt{STELLA} model much better recovers the early optical light curve behavior; the optical is very well reproduced from $1.5-6$ days. At similar phases, the best matched \texttt{CMFGEN} model is still rising and underestimates the optical and near-UV flux. Beyond $\sim$6 days, the \texttt{STELLA} model diverges from the observations and far overestimates the plateau flux while the \texttt{CMFGEN} model reproduces the plateau well.

The UV light curve, particularly at early times, is most important as it is likely dominated by flux excess originating from the CSM interaction. In this domain, neither model matches the observations very well, however, the \texttt{STELLA} model better traces the evolution, especially beyond $\sim$4 days after $t_{fl}$. The \texttt{CMFGEN} model overestimates the UV flux across all bands and all epochs of UV photometry. 

\subsection{Physical parameter inference from model matches}
\label{sec:physical_params}

The key progenitor and CSM parameters of interest are the mass-loss rate ($\dot{M}$) and the CSM density profile ($\rho_\mathrm{CSM}$). Both models adopt density profiles which exponentially decline from the surface of the star outwards. The \cite{Moriya+2023} grid parametrizes the CSM structure as $\beta$ and includes models where the density decreases sharply at the surface of the star when transitioning into the CSM (smaller $\beta$) and models where the density decays as a roughly continuous exponential from within the star to beyond $10^{15}$~cm (large $\beta$; see Fig.~1 in \citealt{Moriya+2023}). The best matched \texttt{STELLA} model uses the most gradually declining CSM density profile for the selected progenitor star ZAMS mass $M_\star$, which corresponds to $\beta=5.0$ or $\rho\propto r^{-1.674}$. The \texttt{CMFGEN} model's density profile declines more rapidly ($\rho\propto r^{-3.174}$), indicating that the matter in the CSM is concentrated closer to the star.

The best matched models infer a mass-loss rate consistent with other SNe exhibiting similar spectroscopic features. The best matched \texttt{CMFGEN} model has $\dot{M}=10^{-3}~M_\odot$ yr$^{-1}$ while the best matched \texttt{STELLA} model has $\dot{M}=10^{-4}~M_\odot$ yr$^{-1}$. These values roughly agree with mass-loss rates of SNe in Fig.~\ref{fig:spec_comparison} inferred in \citetalias{Jacobson-Galan+2024_FMII} using \texttt{CMFGEN} models. Furthermore, the difference of an order of magnitude is not concerning because the model grids do not provide fine granularity in values of $\dot{M}$. Based on spectra from \texttt{CMFGEN} models with greater ($\dot{M}\geq10^{-2}~M_\odot$ yr$^{-1}$) or smaller ($\dot{M}\leq10^{-5}~M_\odot$ yr$^{-1}$) mass-loss rates, we can also confidently rule out those ranges of $\dot{M}$. Models with much larger $\dot{M}$ tend to show stronger and additional narrow emission features with greater durations; models with much smaller $\dot{M}$ show little to no narrow emission features.


The duration of enhanced mass-loss ($t_{\dot{M}}$) can be inferred from the shock velocity, duration of IIn-like features, and the wind velocity: $t_{\dot{M}}=v_{sh}t_\mathrm{IIn}/v_w$. To make this inference, we adopt the indirectly inferred $t_\mathrm{IIn}$ from Sec.~\ref{sec:spec} ($t_\mathrm{IIn}=0.96\pm0.34$ d). We also use the expansion velocity measured from the blue-edge of the H$\alpha$ absorption as a lower-limit for the shock velocity, which yields $v_{sh}\geq18,000$ km~s$^{-1}$ from the Keck/LRIS spectrum. This cannot be precisely measured from earlier spectra as they either do not show a broad H$\alpha$ feature, are not of sufficient resolution, or, in the case of the Binospec spectrum, exhibit blending between H$\alpha$ and Si~II. The $t_{\dot{M}}$ also depends on the wind velocity, which varies between either model: $v_w=50$ km~s$^{-1}$ for \texttt{CMFGEN} and $v_w=10$ km~s$^{-1}$ for \texttt{STELLA}. These values produce a lower limit on the duration of enhanced mass-loss: assuming \texttt{CMFGEN} $v_w$, $t_{\dot{M}}\geq3.0\times10^7~\mathrm{s}\approx1$ yr; assuming \texttt{STELLA} $v_w$, $t_{\dot{M}}\geq1.5\times10^8~\mathrm{s}\approx5$ yr. 

Matching to the extremely large \cite{Moriya+2023} grid also provides suggestions for the values of additional progenitor and explosion properties like the mass of $^{56}$Ni ($M_{\mathrm{Ni}}$), the progenitor star ZAMS mass ($M_\star$), and the explosion energy ($E$).
\cite{Dessart_WJG_2023} do not consider varying these parameters as their study is conducted to qualitatively reproduce IIn-like features. 

The best matched \texttt{STELLA} model adopts $M_{\mathrm{Ni}}=0.04~M_\odot$, $M_\star=10~M_\odot$, and $E=1.5\times10^{51}$ erg. Models with higher explosion energies ($E \gtrsim 3\times10^{51}$ erg) tend to better reproduce the rapid rise observed in SN\,2024jlf but are associated with significantly brighter plateaus. Our best-matched model represents a compromise, balancing the need to match both the rise time and the plateau brightness. 

The $M_\star$ of the best matched model, which controls the duration of the plateau, is the smallest value considered in the grid. The progenitor models adopted from \citet{Sukhbold+2016} by the \cite{Moriya+2023} grid couple the progenitor mass and radius. Thus, this is not a conclusive inference because there are degeneracies between ejecta mass, radius, and explosion energy \citep{Goldberg+2019}. These inferred properties should be interpreted as one plausible solution, rather than a unique configuration.


Lastly, we reiterate that none of the models in either grid reproduce all observed properties of SN\,2024jlf. Both best matched models underestimate the earliest epochs of optical and near-UV photometry by $\sim$1 mag. 

\subsection{The future of \texttt{BTSbot-nearby} and rapid, autonomous follow-up}

\texttt{BTSbot-nearby} is the latest in a rich history of rapid transient follow-up efforts conducted following discoveries by the Palomar 48-inch telescope. Whether for infant SNe or gamma ray bursts, collecting valuable data shortly after transient discovery has long been a core science goal of these programs \citep{Cenko+2006, Graham+2019}. Most recently and most closely related to \texttt{BTSbot-nearby}, auto-triggering technology was developed in the \texttt{AMPEL} broker \citep{Nordin+2019} with \texttt{SNGuess} \citep{Miranda+2022} being used to select targets for follow-up. Automation of transient identification and follow-up, as demonstrated with \texttt{BTSbot-nearby}, is a necessary advancement to previous efforts to minimize latency and collect the earliest possible data. 

Further minimization of latency from the current state of \texttt{BTSbot-nearby} would be very challenging. The limited sensitivity of ZTF aside, one of the greatest remaining sources of latency is the requirement of two detections before triggering follow-up to reject moving objects. The ZTF cadence and scheduling tends to make the latency associated with this criterion $\gtrsim1$ hour. If the \texttt{BTSbot-nearby} filtering were improved to be able to reject moving objects (i.e., asteroids and satellites), the typical latency could be reduced. Section~5.3 of \cite{Rehemtulla+2024_BTSbot} illustrates that a program of this sort could have expedited the follow-up of SN\,2023ixf by $\sim$10 hours. Nevertheless, such aggressive follow-up remains impossible in the vast majority of cases.

\texttt{BTSbot-nearby} now regularly triggers target-of-opportunity (ToO) requests to SEDM and aids in the collection of very early follow-up data. In the case of SN\,2025ay, \texttt{BTSbot-nearby} discovered the transient \citep{25ay_discovery} and collected a spectrum with SEDM just 23 minutes later. The scientific value of the data, irrespective of how early it may be taken, is limited by the data quality. This was a major challenge in \citetalias{Bruch+2023}, where SEDM data often left it ambiguous whether or not narrow \ion{He}{2} emission lines were present. Another future advancement for \texttt{BTSbot-nearby} lies in triggering larger telescopes equipped with more capable instruments. \texttt{BTSbot-nearby} is 
already capable of triggering new urgency 0 ToOs to \textit{Swift}/UVOT \citep{Tohuvavohu+2024} as well as triggers to the 4.1-m Sourthern Astrophysical Research (SOAR) telescope through the Astronomical Event Observatory Network \citep[AEON;][]{Street+2020} integration in Fritz/SkyPortal.

Automation of transient workflows can also act as a service to the community. In the end-to-end automated BTS workflow \citep{Rehemtulla_2023tyk}, new transients are reported to TNS by \texttt{BTSbot} and classifications are reported by \texttt{SNIascore} \citep{Fremling+2021} and \texttt{pySEDM} \citep{Rigault+2019}. Making these data and findings available publicly and quickly enables timely follow-up by others in the community. This is being expanded to CCSNe with the use of \texttt{CCSNscore} \citep{Sharma+2024}. Moreover, this practice synergizes very well with \texttt{BTSbot-nearby}: very early spectra made publicly available immediately can then motivate follow-up with larger facilities by anyone in the community.

\section{Summary} \label{sec:summary}

We have presented the new \texttt{BTSbot-nearby} program, which discovered the 18.5 Mpc Type IIP SN\,2024jlf and obtained spectroscopic follow-up just $+$0.7 days after first light (Sec.~\ref{sec:discovery}). \texttt{BTSbot-nearby} autonomously triggers ToOs for new transients identified by the \texttt{BTSbot} model that are coincident with nearby ($D<60$ Mpc) galaxies (Sec.~\ref{sec:btsbot_nearby}). 

The early spectra of SN\,2024jlf reveal flash ionization features in H$\alpha$, \ion{C}{4}, and \ion{He}{2}, which persist for $1.3<\tau\mathrm{\,[d]}<1.8$ (Sec.~\ref{sec:spec}). With deep non-detections shortly prior to first light, we find that SN\,2024jlf rises exceptionally rapidly, quicker than 90\% of SNe\,II in a large ZTF sample (Sec.~\ref{sec:phot}).
SN\,2024jlf later appears as a normal SN\,IIP with an $\sim$85 day plateau phase and broad, prominent Balmer P-Cygni features.

We match our observations to model grids produced by two independent radiation hydrodynamics codes. The best matched \texttt{CMFGEN} model well reproduces our spectral series and the plateau phase optical brightness although it systematically overestimates the UV flux (Sec.~\ref{sec:cmfgen_models}). The best matched \texttt{STELLA} model much better reproduces the early and UV photometry despite dramatically overestimating the plateau phase optical brightness (Sec.~\ref{sec:moriya_models}). Either model is associated with physical parameters, which can provide suggestions to the nature of SN\,2024jlf's RSG progenitor (Sec.~\ref{sec:discussion}). The mass-loss rates inferred from either model are roughly consistent: $\dot{M}=10^{-3}~M_\odot$ yr$^{-1}$ from the \texttt{CMFGEN} model and $\dot{M}=10^{-4}~M_\odot$ yr$^{-1}$ from the \texttt{STELLA} model. Moreover, models with larger or smaller $\dot{M}$ are clearly disfavored in both grids, so we infer $10^{-4}~<~\dot{M}~\mathrm{[M_\odot~yr^{-1}]}~<~10^{-3}$. Based on adopted values of wind velocity by either model, they suggest the enhanced mass-loss phase persists for $1~<~t_{\dot{M}}\,[\mathrm{yr}]~<~5$. 

These findings demonstrate the value of automated ToO follow-up of transients for probing the nature of flash ionization in CCSNe. Automating first-response follow-up to the initial discovery of a young transient can provide the opportunity to collect data at otherwise inaccessible phases of the transient's evolution.





\section{Acknowledgments}

W. M. Keck Observatory, MMT Observatory, and Zwicky Transient Facility access for N.R., A.A.M., S.S., and C.L. was supported by Northwestern University and the Center for Interdisciplinary Exploration and Research in Astrophysics (CIERA).

The material contained in this document is based upon work supported by a National Aeronautics and Space Administration (NASA) grant or cooperative agreement. Any opinions, findings, conclusions, or recommendations expressed in this material are those of the author and do not necessarily reflect the views of NASA. This work was supported through a NASA grant awarded to the Illinois/NASA Space Grant Consortium.

This research was supported in part through the computational resources and staff contributions provided for the Quest high performance computing facility at Northwestern University which is jointly supported by the Office of the Provost, the Office for Research, and Northwestern University Information Technology.

Based on observations obtained with the Samuel Oschin Telescope 48-inch and the 60-inch Telescope at the Palomar Observatory as part of the Zwicky Transient Facility project. ZTF is supported by the National Science Foundation under Grants No. AST-1440341 and AST-2034437 and a collaboration including current partners Caltech, IPAC, the Oskar Klein Center at Stockholm University, the University of Maryland, University of California, Berkeley, the University of Wisconsin at Milwaukee, University of Warwick, Ruhr University, Cornell University, Northwestern University and Drexel University. Operations are conducted by COO, IPAC, and UW.

The Young Supernova Experiment is supported by the National Science Foundation through grants AST-1518052, AST-1815935, AST-1852393, AST-1911206, AST-1909796, and AST-1944985; the David and Lucile Packard Foundation; the Gordon \& Betty Moore Foundation; the Heising-Simons Foundation; NASA through grants NNG17PX03C, 80NSSC19K1386, and 80NSSC20K0953; the Danish National Research Foundation through grant DNRF132; VILLUM FONDEN Investigator grants 16599, 10123 and 25501; the Science and Technology Facilities Council through grants ST/P000312/1, ST/S006109/1 and ST/T000198/1; the Australian Research Council Centre of Excellence for All Sky Astrophysics in 3 Dimensions (ASTRO 3D) through project number CE170100013; the Hong Kong government through GRF grant HKU27305119; the Independent Research Fund Denmark via grant numbers DFF 4002-00275 and 8021-00130, and the European Union’s Horizon 2020 research and innovation programme under the Marie Sklodowska-Curie through grant No. 891744.

The Pan-STARRS1 Surveys (PS1) and the PS1 public science archive have been made possible through contributions by the Institute for Astronomy, the University of Hawaii, the Pan-STARRS Project Office, the Max-Planck Society and its participating institutes, the Max Planck Institute for Astronomy, Heidelberg and the Max Planck Institute for Extraterrestrial Physics, Garching, The Johns Hopkins University, Durham University, the University of Edinburgh, the Queen’s University Belfast, the Harvard-Smithsonian Center for Astrophysics, the Las Cumbres Observatory Global Telescope Network Incorporated, the National Central University of Taiwan, the Space Telescope Science Institute, the National Aeronautics and Space Administration under Grant No. NNX08AR22G issued through the Planetary Science Division of the NASA Science Mission Directorate, the National Science Foundation Grant No. AST-1238877, the University of Maryland, Eotvos Lorand University (ELTE), the Los Alamos National Laboratory, and the Gordon and Betty Moore Foundation.

YSE computations are aided by the University of Chicago Research Computing Center, the Illinois Campus Cluster, and facilities at the National Center for Supercomputing Applications at UIUC.

Observations reported here were obtained at the MMT Observatory, a joint facility of the University of Arizona and the Smithsonian Institution. 

The authors wish to recognize and acknowledge the very significant cultural role and reverence that the summit of Mauna Kea has always had within the indigenous Hawaiian community. We are most fortunate to have the opportunity to conduct observations from this mountain. 

SED Machine is based upon work supported by the National Science Foundation under Grant No. 1106171. The ZTF forced-photometry service was funded under the Heising-Simons Foundation grant \#12540303 (PI: Graham). The Gordon and Betty Moore Foundation, through both the Data-Driven Investigator Program and a dedicated grant, provided critical funding for SkyPortal.

Based on observations made with the Nordic Optical Telescope, owned in collaboration by the University of Turku and Aarhus University, and operated jointly by Aarhus University, the University of Turku and the University of Oslo, representing Denmark, Finland and Norway, the University of Iceland and Stockholm University at the Observatorio del Roque de los Muchachos, La Palma, Spain, of the Instituto de Astrofisica de Canarias.

N.R., C.L.,~and A.A.M.~are supported by DoE award \#DE-SC0025599. W.J-G. is supported by NASA through the NASA Hubble Fellowship grant HSTHF2-51558.001-A awarded by the Space Telescope Science Institute, which is operated by the Association of Universities for Research in Astronomy, Inc., for NASA, under contract NAS5-26555. Parts of this research were supported by the Australian Research Council Centre of Excellence for Gravitational Wave Discovery (OzGrav), through project number CE230100016 and the Australian Research Council Discovery Early Career Researcher Award (DECRA) through project number DE230101069. M.W.C acknowledges support from the National Science Foundation with grant numbers PHY-2308862 and PHY-2117997. D.O.J. acknowledges support from NSF grants AST-2407632 and AST-2429450, NASA grant 80NSSC24M0023, and HST/JWST grants HST-GO-17128.028, HST-GO-16269.012, and JWST-GO-05324.031, awarded by the Space Telescope Science Institute (STScI), which is operated by the Association of Universities for Research in Astronomy, Inc., for NASA, under contract NAS5-26555. S.R.K. thanks the Heising-Simons Foundation for supporting his research. 

This research has made use of NASA's Astrophysics Data System.

Software citation information aggregated using The Software Citation Station \citep{software-citation-station-paper, software-citation-station-zenodo}.


%

\vspace{5mm}
\facilities{PO:1.2m (ZTF), PO:1.5m (SEDM), LT (SPRAT), MMT (Binospec), NOT (ALFOSC), Keck:I (LRIS), Keck:II (KCWI), ATT (WiFeS)}


\software{
    \texttt{astropy} \citep{astropy:2013, astropy:2018, astropy:2022}, \texttt{Jupyter} \citep{IPython, Jupyter}, \texttt{Keras} \citep{keras}, \texttt{matplotlib} \citep{Matplotlib}, \cite{NED}, \texttt{numpy} \citep{Numpy}, \texttt{pandas} \citep{Scipy_2010, pandas_2024}, \texttt{penquins} \citep{penquins}, \texttt{python} \citep{Python3}, \texttt{scipy} \citep{Scipy_2010, SciPy_2020, Scipy_2024}, \texttt{astroquery} \citep{Astroquery_2019, Astroquery_2024}, \texttt{Cython} \citep{cython:2011}, \texttt{dustmaps} \citep{dustmaps_2018, dustmaps_2024}, \texttt{h5py} \citep{dustmaps_2018, dustmaps_2024}, \texttt{SkyPortal} \citep{van_der_Walt+2019, Coughlin+2023}, \texttt{tensorflow} \citep{tensorflow}, \texttt{tqdm} \citep{tqdm}, and the Weights and Biases platform \citep{wandb}. The \texttt{BTSbot-nearby} alert filter in MongoDB aggregation pipeline syntax and additional utilities for testing it can be found at \url{https://github.com/nabeelre/BTSbot-nearby-utils}.
}


\appendix

\section{Disentangling Flash Features from the Host Emission} \label{app:flash_2d}
\begin{figure}
    \centering
    \includegraphics[width=0.75\linewidth]{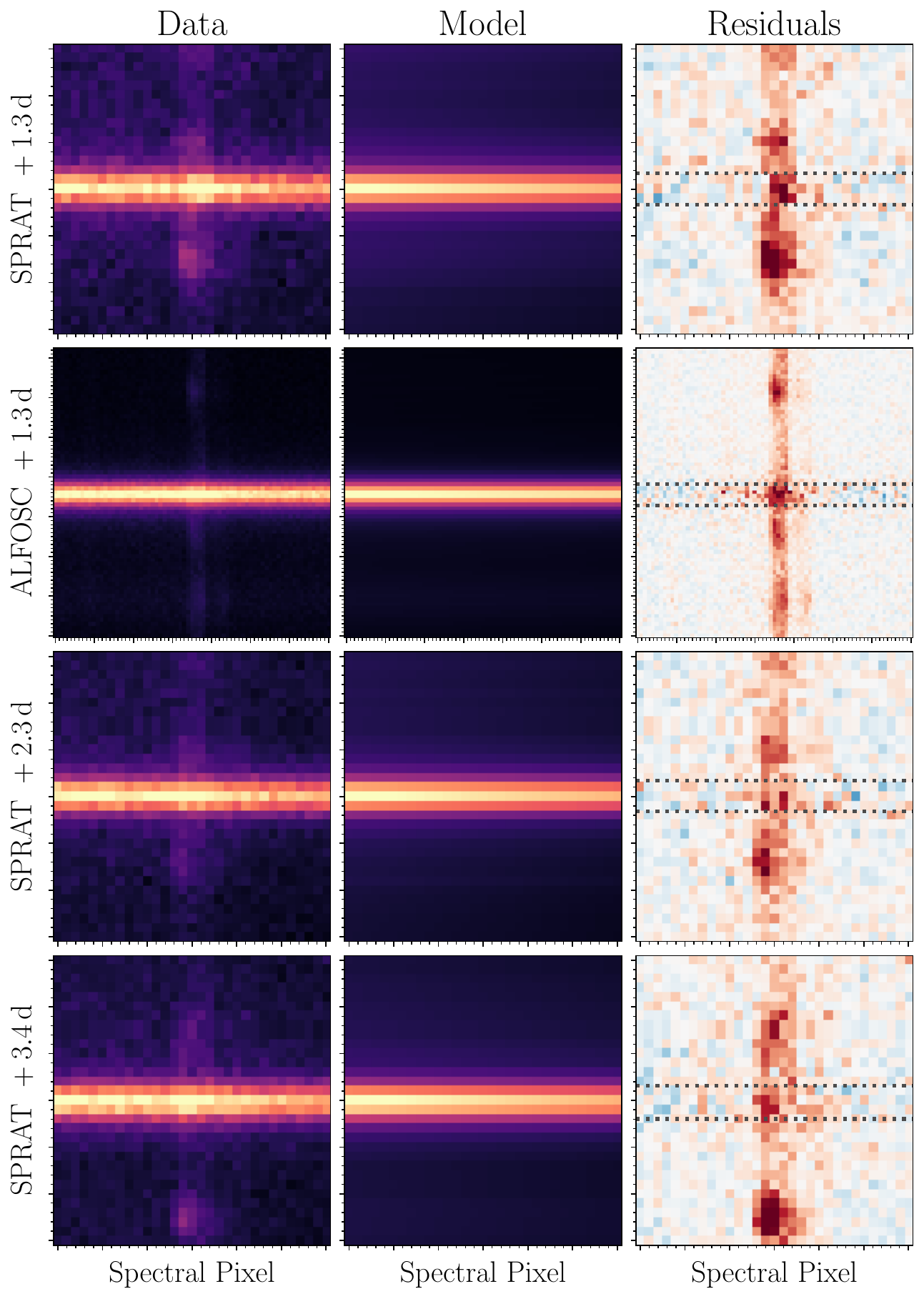}
    \caption{The disappearance of the H$\alpha$ flash feature viewed from early 2D spectra. \textit{Left column:} the calibrated 2D spectra in the vicinity of the H$\alpha$ emission; \textit{Center column:} models of the SN continua and the global background, including the host galaxy continua and the sky background (see the text for modeling details); \textit{Right column:} flux residuals originating from narrow SN and host emission lines, with the FWHM of the SN trace overlaid (dotted lines). Each cutout spans $\sim$150\,\r{A} in spectral direction (centered at 6600\,\r{A}) and $\sim$14'' in spatial direction. The host galaxy is spatially resolved along the slit, and the vertical stripes in the left and right columns correspond to H$\alpha$ emission from star-forming regions covered by the slit. Each observation was made with the slit placed at a specific, different position angle; thus, the H$\alpha$ patterns along the slit direction vary between observations. At +1.3\,d, there is residual flux centered within the FWHM of the SN but shifted $\sim$1\,pixel rightward of the host emission's centroid. This indicates the existence of a slightly redshifted ($\sim$$10^2$\,km\,s$^{-1}$) component intrinsic to the SN. There is no evidence for such a component in the +2.3\,d and +3.4\,d spectra, indicating the disappearance of narrow H$\alpha$ emission from the SN.}
    \label{fig:flash_2d}
\end{figure}

Emission lines (e.g., H$\alpha$) from the host galaxy can contaminate the SN spectrum, and it is critical (and often not trivial) to disentangle flash features of the SN from the host emission. To precisely determine when the H$\alpha$ flash feature disappears, in Figure~\ref{fig:flash_2d} we present the 2D spectra obtained before +3.4\,d with LT/SPRAT and NOT/ALFOSC, which encode the spatial distribution of H$\alpha$ emission from the host. The calibrated 2D spectra all exhibit a vertical stripe that overlaps the horizontal trace of the SN continuum, corresponding to the H$\alpha$ emission from star-forming regions in the host galaxy covered by the slit. Because the slit was oriented at a different position angle for each observations, the H$\alpha$ background varies over observations. In the vicinity of the H$\alpha$ features ($\sim$150\,\r{A} as displayed in the image cutouts), we assume (i) that the flux density of the SN continuum can be approximated with a linear function of wavelength, and (ii) that the PSF of the SN remains the same at different wavelengths. The contribution of the SN continuum at pixel $(x_\mathrm{spec}, x_\mathrm{spat})$ is
\begin{equation}
    F_\mathrm{SN, con}(x_\mathrm{spec}, x_\mathrm{spat}) = \left(k_\mathrm{SN}x_\mathrm{spec} + b_\mathrm{SN}\right)\cdot \mathrm{PSF}(x_\mathrm{spat}).
\end{equation}
We estimate the PSF using data outside a $l_\mathrm{mask}=15$ pixel mask covering the spectral pixels contaminated by the H$\alpha$ emission. At each $x_\mathrm{spat}$ we have
\begin{equation}
    \mathrm{PSF}(x_\mathrm{spat}) = \underset{{|x_\mathrm{spec} - x_{\mathrm H\alpha}| > l_\mathrm{mask}/2}}{\mathrm{Median}}\left\{\frac{F_\mathrm{obs}(x_\mathrm{spec}, x_\mathrm{spat}) - F_\mathrm{sky}(x_\mathrm{spec}, x_\mathrm{spat})}{\sum_{x_\mathrm{spat}} \left[F_\mathrm{obs}(x_\mathrm{spec}, x_\mathrm{spat}) - F_\mathrm{sky}(x_\mathrm{spec}, x_\mathrm{spat})\right]}\right\}.
\end{equation}
To model the sky background, which consists of both the sky emission and the host galaxy continuum, we also assume its linear dependence on wavelength,
\begin{equation}
    F_\mathrm{sky}(x_\mathrm{spec}, x_\mathrm{spat}) = k_\mathrm{sky}x_\mathrm{spec} + b_\mathrm{sky}.
\end{equation}
Outside the H$\alpha$ region, the sum of $F_\mathrm{SN,con}$ and $F_\mathrm{sky}$ should be a good representation of the total flux. To find the optimized model parameters ($k_\mathrm{SN}$, $b_\mathrm{SN}$, $k_\mathrm{sky}$, $b_\mathrm{sky}$), we minimize the residual sum of squares outside the H$\alpha$ mask. The resultant models are presented in the center column of panels in Figure~\ref{fig:flash_2d}.

The right panels of Figure~\ref{fig:flash_2d} show the flux residuals, which should correspond to the narrow H$\alpha$ emission of both the host galaxy and the SN. In each panel, we overlay the FWHM of the SN trace as dotted lines. In the +1.3\,d LT/SPRAT spectrum, there is a blob of emission features right at the center of the SN trace (see also Figure~\ref{fig:flash_features}), whose centroid is $\sim$1 pixel rightward of the centroid of the host emission. This indicates the existence of another H$\alpha$ component that is slightly ($\sim$$10^2$\,km\,s$^{-1}$) redshifted relative to the galaxy background, which is consistent with a flash feature of the SN. In the NOT/ALFOSC spectrum obtained at nearly the same time, the emission emerges at the consistent wavelength. There is no evidence of such an offset component in the +2.3\,d and +3.4\,d spectra, confirming that the H$\alpha$ flash feature appears to be absent by +2.3\,d. A narrow H$\alpha$ feature is visible in the +3.4\,d 1D spectrum (Figure~\ref{fig:flash_features}), but the location of this emission feature in the 2D spectrum is consistent with the centroid of the host emission. Thus, the H$\alpha$ feature in the 1D spectrum is contamination from the host and not a flash feature originating from the SN.


\clearpage
\bibliography{main}{}
\bibliographystyle{aasjournal}


\end{CJK*}
\end{document}